\documentclass[twocolumn]{aastex631}
\let\tablenum\relax
\usepackage{siunitx}
\usepackage{multirow}
\usepackage{tabularx}
\shorttitle{Modeling of the high-velocity jet powered by the massive star MWC 349A}
\shortauthors{Martínez-Henares et al.}
\graphicspath{{./}{figures/}}

\begin{document}

\title{Modeling of the high-velocity jet powered by the massive star MWC 349A}

\author[0000-0001-5191-2075]{Antonio Martínez-Henares}
\affil{Centro de Astrobiología (CSIC-INTA)\\
Ctra de Torrejón a Ajalvir, km 4 \\
28850 Torrejón de Ardoz, Madrid, Spain}

\author[0000-0003-4493-8714]{Izaskun Jiménez-Serra}
\affil{Centro de Astrobiología (CSIC-INTA)\\
Ctra de Torrejón a Ajalvir, km 4 \\
28850 Torrejón de Ardoz, Madrid, Spain}

\author[0000-0003-4561-3508]{Jesús Martín-Pintado}
\affil{Centro de Astrobiología (CSIC-INTA)\\
Ctra de Torrejón a Ajalvir, km 4 \\
28850 Torrejón de Ardoz, Madrid, Spain}

\author[0000-0002-2711-8143]{Nuria Huélamo}
\affil{Centro de Astrobiología (CSIC-INTA)\\
Ctra de Torrejón a Ajalvir, km 4 \\
28850 Torrejón de Ardoz, Madrid, Spain}

\author[0000-0002-1082-5589]{Sirina Prasad}
\affil{Center for Astrophysics $\vert$ Harvard \& Smithsonian \\
60 Garden Street \\
Cambridge, MA 02138, USA}

\author[0000-0003-2384-6589]{Qizhou Zhang}
\affil{Center for Astrophysics $\vert$ Harvard \& Smithsonian \\
60 Garden Street \\
Cambridge, MA 02138, USA}

\author[0000-0002-3882-4414]{James Moran}
\affil{Center for Astrophysics $\vert$ Harvard \& Smithsonian \\
60 Garden Street \\
Cambridge, MA 02138, USA}

\author[0000-0002-6368-7570]{Yue Cao}
\affil{Center for Astrophysics $\vert$ Harvard \& Smithsonian \\
60 Garden Street \\
Cambridge, MA 02138, USA}

\author{Alejandro Báez-Rubio}
\affil{UWC Mahindra College \\
Village Khubavali, PO Paud \\
MH 412 108, India}



\begin{abstract}

MWC 349A is a massive star with a well-known circumstellar disk rotating following a Keplerian law, and an ionized wind launched from the disk surface. Recent ALMA observations carried out toward this system have however revealed an additional high-velocity component in the strong, maser emission of hydrogen radio recombination lines (RRLs), suggesting the presence of a high-velocity ionized jet. In this work, we present 3D non-LTE radiative transfer modeling of the emission of the H30$\alpha$ and H26$\alpha$ maser lines, and of their associated radio continuum emission, toward the MWC 349A massive star. By using the MORELI code, we reproduce the spatial distribution and kinematics of the high-velocity emission of the H30$\alpha$ and H26$\alpha$ maser lines with a high-velocity ionized jet expanding at a velocity of $\sim$ 250 \unit{km.s^{-1}}, surrounded by MWC 349A's wide-angle ionized wind. The bipolar jet, which is launched from MWC 349A's disk, is poorly collimated and slightly miss-aligned with respect to the disk rotation axis. Thanks to the unprecedented sensitivity and spatial accuracy provided by ALMA, we also find that the already known, wide-angle ionized wind decelerates as it expands radially from the ionized disk. We briefly discuss the implications of our findings in understanding the formation and evolution of massive stars. Our results show the huge potential of RRL masers as powerful probes of the innermost ionized regions around massive stars and of their high-velocity jets.

\end{abstract}



\section{Introduction} \label{sec:intro}

    In the past decade, it has been proposed that isolated massive stars (with masses $\geq8$ M$_{\odot}$) grow their mass via gas accretion through circumstellar disks \citep{mckeetan2003, rosenkrumholz2020} which can last even during the main sequence phase of their evolution \citep{keto2003}. These accreting systems host expanding outflows that remove the excess of angular momentum in a similar fashion to low-mass stars (see \citealt{frank2014} for a review). In the late stages of their formation, massive stars emit vast amounts of UV radiation that strongly ionize their surroundings, causing the photoevaporation of their circumstellar disk \citep{hollenbach1994,gortihollenbach2009}, and forming ultracompact (UC) HII regions \citep[e.g.][] {garaylizano1999,keto2008}.
    
    A significant fraction ($\sim$30\%) of the observed UCHII regions emit broad (70-200 \unit{km.s^{-1}}) hydrogen radio recombination lines (RRLs) and have rising power law continuum spectra with spectral indices $\alpha\sim$ 0.6-1.5, which are interpreted to arise from ionized winds formed in neutral disks \citep{reynolds1986,jaffemartinpintado1999}. Hydrogen RRLs that undertake maser amplification allow us to perform spectro-astrometry studies reaching an excellent photocenter accuracy down to $\sim$1 mas or better, which would be otherwise unachievable. Some well-known examples of RRL maser emission in massive stars are MWC 349A \citep{martinpintado1989, planesas1992,martinpintado2011,baezrubio2013}, Cep A HW2 \citep{jimenezserra2011}, Mon R2-IRS2 \citep{jimenezserra2013,jimenezserra2020}, $\eta$ Carinae \citep{abraham2014}, MWC922 \citep{sanchezcontreras2019} or G45.47+0.05 \citep{zhang2019}. In the latter, the authors find hints of a jet embedded within the wide-angle photoevaporative outflow through the detection of a jet-like structure with negative spectral index between 7mm and 1.3mm. Cep A HW2 presents a photoevaporating rotating disk perpendicular to a high-velocity collimated jet \citep{jimenezserra2007} within a slow wide-angle outflow probed by water masers expanding simultaneously with the jet \citep{torrelles2011}.
    
    A more recent example of a massive star with an ionized disk plus a wide-angle wind and a high-velocity jet is provided by the massive star MWC 349A. The star, located at a distance of \mbox{1.2 kpc} \citep{cohen1985}, is surrounded by a circumstellar disk \citep{danchi2001} that rotates following a Keplerian law \citep{planesas1992,weintroub2008}. A wide-angle wind is launched from the disk \citep[see e.g.][]{martinpintado2011,baezrubio2014,zhang2017} at a nearly constant velocity of \mbox{$\sim$60 km s$^{-1}$} \citep{olnon1975,tafoya2004}. Observations of RRLs with the PdBI \citep{martinpintado2011,baezrubio2013} and the SMA \citep{zhang2017} have shown that the outflow is rotating in the same sense as the disk, extracting angular momentum from the system. 
    
    In addition to the ionized disk and expanding wide-angle wind, \cite{prasad2023} have recently reported the presence of an additional velocity component in the emission of the bright H30$\alpha$ and H26$\alpha$ RRL maser emission toward MWC 349A at radial velocities from \mbox{$-$85 to $-$40 km s$^{-1}$} and from \mbox{+60 to +100 km s$^{-1}$}, which had never been explored before. The 2D Gaussian centroid positions associated with the high-velocity RRL maser component appear in a linear structure nearly aligned with the polar axis of the disk and thus, \cite{prasad2023} have postulated that this component may be associated with a high-velocity jet magneto-hydrodynamically launched from the disk. If these masers lied close to the polar axis of the system, the velocities of the high-velocity jet could be as high as 575 km s$^{-1}$. However, the detailed physical structure and kinematics of this high-velocity jet in MWC 349A remain unknown. 
    
    To characterize the jet in MWC 349A, we have carried out a detailed 3D non-LTE radiative transfer modeling of the RRL maser emission toward this source, recently measured with ALMA in \cite{prasad2023}. Thanks to the exquisite precision of the 2D Gaussian centroid positions obtained with ALMA, our model provides the structure, orientation, expanding velocity and rotation of the high-velocity jet embedded within the already known ionized wind in MWC 349A. The model is described in Sect. \ref{sec:model} and its results are presented in Sect. \ref{sec:results}. We discuss these results in view of theories of massive star formation in Sect. \ref{sec:discussion} before presenting our conclusions in Sect. \ref{sec:conclusions}.

\section{Non-LTE radiative transfer model} \label{sec:model}

    We have carried out the radiative transfer modeling of the H30$\alpha$ and H26$\alpha$ lines for MWC 349A using the MORELI code \citep{baezrubio2013}. MORELI stands for MOdel for REcombination LInes and it is a non-LTE 3D radiative transfer code that models the emission from RRLs and their adjacent radio continuum arising from ionized regions with rotational symmetry. The ionized region to be modeled is discretized in a 3D mesh of regular cubes with sizes $(dx,dy,dz)$. The $z$ axis is taken to be along the line-of-sight, with negative values towards the observer; the $x$ axis corresponds to the projection of the revolution axis of the source on the plane of the sky; and the $y$ axis is orthogonal to the $x$ axis in the plane of the sky. The radiative transfer equation is integrated along all lines of sight.
    
    In previous works, the kinematics of the ionized gas in MWC 349A has been modeled considering only two components: an ionized rotating disk and a wide-angle ionized bipolar outflow in a double-cone geometry of opening angle $\theta_a$. The ionized disk is the boundary layer of opening angle $\theta_d$ (Figure \ref{fig:sketch}) ionized by the central star, located between a neutral disk and the ionized outflow originated by the photoevaporation of the disk. The ionized gas of the disk is modeled using an electronic density $N_{e,d}$ and electronic temperature $T_d$. The ionized wind is described using a double cone geometry with an opening angle $\theta_w=\theta_a-\theta_d$, internal to the ionized disk, and electronic density distribution $N_{e,w}$ and electronic temperature $T_0$. The electron density distribution of the ionized material in the double cone depends on the distance to the center $r$ and on the angle with respect to the revolution axis $\theta$ as \citep{baezrubio2013}:

    \begin{equation}
        N_e(r,\theta)=N_{e,0}\frac{e^{-(\theta_a-\theta)/\theta_0}}{r^{b_d}}
    \label{eq:distribution}
    \end{equation}
    
    \noindent where $N_{e,0}$ is a reference value of the density at a reference distance $r_0$ and at $\theta=\theta_a$. $\theta_0$ and $b_d$ are dimensionless parameters that regulate the angular and radial dependence of the distribution, respectively. The model includes the possibility of considering different electron density distributions and/or electron temperatures between the ionized disk and the ionized wind.
    
    The ionized disk rotates around the central star following a Keplerian law \citep{planesas1992,weintroub2008,martinpintado2011}. The wind, in contrast, expands radially from the central star, and can rotate following the Keplerian motion of the disk \citep{baezrubio2013,zhang2017}. However, as shown in Section \ref{sec:results}, this model cannot fully reproduce the newly observed kinematical structure revealed by ALMA in the H30$\alpha$ and in H26$\alpha$ lines. As suggested by \cite{prasad2023}, this component may be associated with a high-velocity jet and therefore, in the following sections, we include in the model a new kinematic component in the ionized outflow of MWC 349A describing the jet. In addition, the comparison with the new ALMA images shows that a deceleration of the radial expansion of the outflow is also required in order to explain the observations. The details on how these new features are introduced in MORELI are reported in Sections \ref{sec:jet} and \ref{sec:bv}, respectively.
    
    \subsection{Jet in the outflow kinematics} \label{sec:jet}
    
\begin{figure*}
            \includegraphics[width=18cm, height=7.8cm]{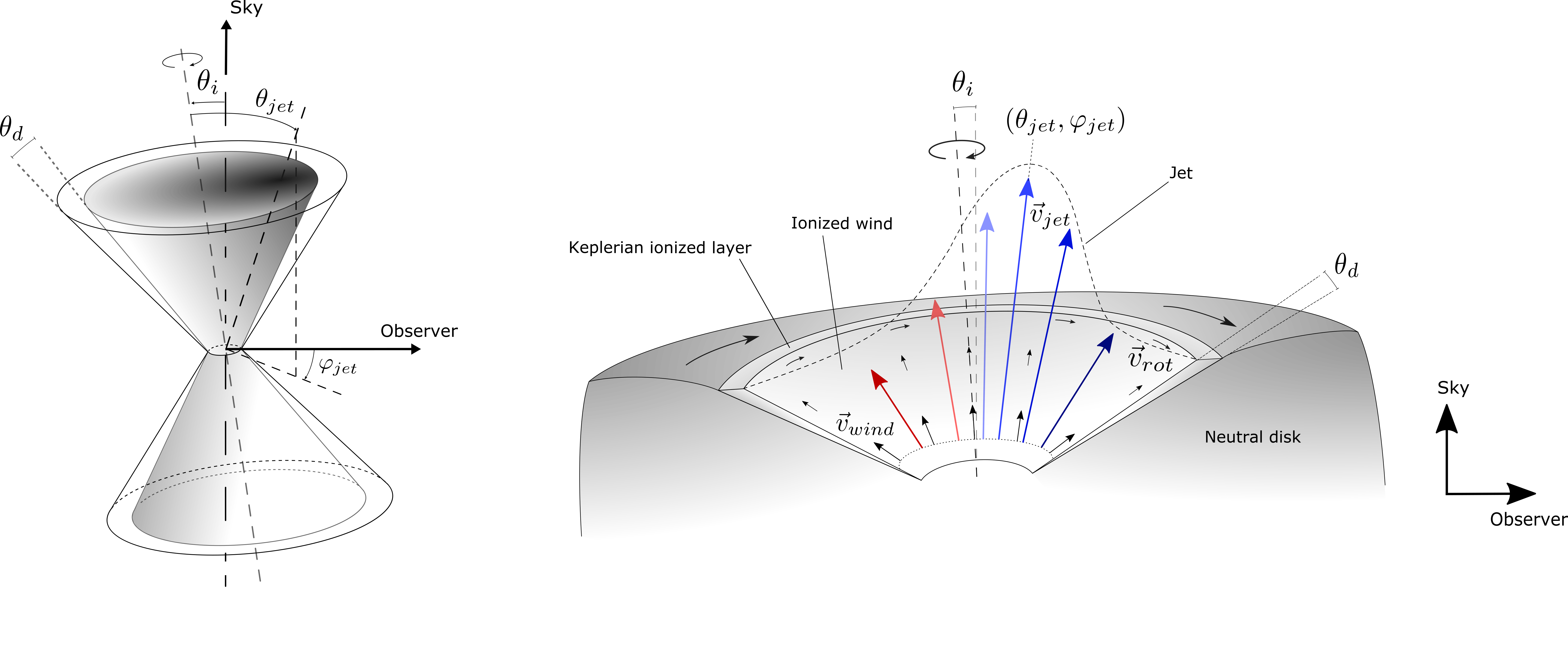}
            \centering
            \caption{Sketches that illustrate the geometry and kinematics of the model. The two figures share common features: the ionized gas is located inside the cone delimited by the layer of thickness $\theta_d$, corresponding to the Keplerian ionized disk. The neutral disk is located beyond this layer. The disk is tilted an angle $\theta_i$ with respect to the plane of the sky; and the jet orientation is represented as ($\theta_{jet}, \varphi_{jet}$). \textit{Left: }Shaded grey areas inside the double cone represent the jet velocity, with darker tones indicating higher velocity values. The jet has a certain collimation, hence the velocity decreases with the angular distance to the jet axis ($\theta_{jet}, \varphi_{jet}$). \textit{Right: } Kinematic components of the model. We show a vertical cut of the northern cone. The dotted line at small radius from the central star indicates the launching radius of the ouflow at $\sim$24 AU of the star \citep{baezrubio2014}. Large colored arrows surrounded by the dotted curved line correspond to the jet expansion velocity $\vec{v}_{jet}$, which becomes slower with larger angular distances from the jet orientation due to collimation. Darker blue (red) colors indicate a higher projection of the blueshifted (redshifted) jet expansion along the line of sight. Straight black arrows represent the wind expansion velocity $\vec{v}_{wind}$ and are smaller at larger radial distances to account for the radial deceleration $b_v$; curved black arrows represent the wind rotation velocity component $\vec{v}_{rot}$ and are smaller in the ionized wind than in the disk, as in the former the wind rotates with half Keplerian velocity in our model (Sect. \ref{sec:centroidmap}). Neither of the figures are to scale. \label{fig:sketch}}
        \end{figure*}

        The jet component consists of an oriented high velocity radial expansion outwards from the central star. To characterize the orientation of the jet, we have defined the polar and azimuthal angles $(\theta,\varphi)$ in the spherical set of coordinates (see Figure \ref{fig:sketch}). The polar angle $\theta$ has a value of $0^{\circ}$ at the rotation axis of the source in the northern half of the disk plane. The azimuthal angle $\varphi$ is defined as the angle with value $0^{\circ}$ at the negative $z$ axis, having positive values clockwise as seen from the northern half of the disk plane. An asymmetrical jet can be introduced by specifying different orientations of the northern and southern lobe.
        
        With these angular coordinates, in equation (\ref{eq:psi}) we define the angular distance $\psi$ from any given direction $(\theta,\varphi)$ to the jet orientation $(\theta_{jet},\varphi_{jet})$ with spherical trigonometry:
        
        \begin{equation}
            \cos{\psi} = \cos{\theta_{jet}}\cos{\theta}+\sin{\theta_{jet}}\sin{\theta}\cos{(\varphi-\varphi_{jet})}
        \label{eq:psi}
        \end{equation}
    
        The definition of this angular distance leads us to introduce the collimation of the jet, which we formulate as an angular weighting of the contribution of the jet to the outflow velocity. To account for any orientation of the jet other than along the rotation axis, we have adapted the formulation of \cite{matznermckee1999} by normalizing the distribution and defining it with the angle $\psi$ rather than $\theta$, as follows:
        
        \begin{equation}
            f(\psi) = \frac{\psi_0^2}{\sin^2{\psi}+\psi_0^2}
        \label{eq:psiweight}
        \end{equation}
        
        The parameter $\psi_0$ acts as a flattening factor for the distribution, i.e. it regulates the degree of collimation of the outflow. Larger angles $\psi$ with respect to the orientation of the jet will have a smaller contribution. Smaller values of $\psi_0$ give more collimated outflows, i.e. the velocity decreases rapidly with the angle from the peak value at the center.
        
        Therefore, the velocity of the expanding jet is described by Equation \ref{eq:jetvel_vector} as:
        
        \begin{equation}
            \mathbf{v_{jet}} = v_{jet} f(\psi) \mathbf{e_r}
        \label{eq:jetvel_vector}
        \end{equation}
        
        \noindent where $v_{jet}$ is the peak velocity at the jet orientation (where $f(\psi)=1)$ and $\mathbf{e_r}$ is the radial unit vector in spherical coordinates.

    \subsection{Deceleration of the outflow expansion} \label{sec:bv}
    
        In addition to the jet, the outflow has a wind component that expands radially away from the central star at a terminal velocity $v_{wind}$ as described in Equation \ref{eq:wind_vector}:
        
        \begin{equation}
            \mathbf{v_{wind}} = v_{wind} \mathbf{e_r}
        \label{eq:wind_vector}
        \end{equation}
        
        The projection of both components, jet plus wind, onto the line of sight (i.e. the radial velocity) is expressed in Equation \ref{eq:windandjet} as derived in \cite{baezrubio2013}:
        
        \begin{equation}
            v_{expansion, r} = [v_{wind} + v_{jet} f(\psi)] \frac{z}{r}
        \label{eq:windandjet}
        \end{equation}
        
        \noindent where $z$ is the distance along the line-of-sight (with the observer located toward negative z values) and $r$ is the distance from the central star.
        
        This radial expansion can be decelerated by introducing a dimensionless parameter $b_v$ as shown in Equation \ref{eq:radialexpbv}:
        
        \begin{equation}
            v_{expansion, r} = [v_{wind} + v_{jet} f(\psi)] \left (\frac{r}{r_0}\right)^{b_v} \frac{z}{r}
            \label{eq:radialexpbv}
        \end{equation}
    
        \noindent where $b_v < 0$ for deceleration and $r_0$ is a characteristic length used in the model as the reference unit for all lengths, expressed explicitly here for dimensional consistency.
        
\section{Results} \label{sec:results}

        \begin{table*}[]
        \caption{Input parameters and best-fit values for the final model of MWC 349A. Note that these are not all the input parameters of the model: here we present those that are most relevant for the discussion and/or that are different from those in \cite{baezrubio2013}. The rest of the parameters can be found in this reference. 
        $^a:N_{e,0}=N_e(r=10$ au$,\theta=\theta_a)=1.6\times10^9$\unit{cm^{-3}}.}
        \begin{center}
        \label{tab:inputparameters}
        \begin{tabular}{c c c}
         \hline
         \hline
         Input parameter & Best-fit value & Constrained from \\  
         \hline
         
         Systemic velocity, $v_{sys}$ & 8 \unit{km.s^{-1}} & Line profile\\

         Disk inclination angle, $\theta_i$ & 8$^{\circ}$ (tipped up) & Centroid map\\
         
         Central mass, $M_*$ & 23 M$_{\odot}$ & Line profile \\ 

         Radius of the Keplerian ionized disk, $r_K$ & 46 au & Line profile \\
         
         \hline

         Density distribution, $N_e(r,\theta)^a$ & $1.6\times10^9 \left(\frac{r}{10\text{au}}\right)^{-2.14} e^{\frac{-(\theta_a-\theta)}{20}}$ \unit{cm^{-3}} & Radio-continuum \\

         Double-cone's semi-opening, $\theta_a$ & 57$^{\circ}$ & Radio-continuum \\
         
         Opening angle of the ionized disk, $\theta_d$ & 13$^{\circ}$ & Line profile \\

         Ionized disk's electron temperature, $T_d$ & 9450 K & Line profile \\

         Ionized wind's electron temperature, $T_0$ & 12000 K & Radio-continuum and line profile \\
         
         \hline
         
         Wind terminal velocity, $v_{wind}$ & 60 \unit{km.s^{-1}} & Centroid map and line profile\\
         
         Outflow deceleration parameter, $b_v$ & -0.05 & Centroid map and line profile \\
         
         Jet peak velocity, $v_{jet}$ & 250 \unit{km.s^{-1}} & Centroid map and line profile \\
         
         Jet collimation, $\psi_0$ & 0.4 & Centroid map and line profile \\
         
         Northern jet orientation, $(\theta_{jet, north},\varphi_{jet, north})$ & (16$^{\circ}$, 0$^{\circ}$) & Centroid map \\
         
         Southern jet orientation, $(\theta_{jet, south},\varphi_{jet, south})$ & (158$^{\circ}$, 180$^{\circ}$) & Centroid map \\
         
         \hline

        \end{tabular}
        \end{center}
        \end{table*}

  In order to constrain the physical structure and the kinematics of the ionized gas in MWC 349A, we compare the results of our radiative transfer modeling with the radiocontinuum maps obtained with ALMA at \mbox{217 GHz} and \mbox{342 GHz} (Section \ref{sec:radiocontinuum}), and with the ALMA H30$\alpha$ and H26$\alpha$ centroid maps, line profiles and rotation curves (see Sections \ref{sec:centroidmap}, \ref{sec:h30alineprofile} and \ref{sec:rotationcurves}) presented in \cite{prasad2023}. The H26$\alpha$ line likely suffers from saturation effects and therefore we also explore this phenomenon in Section \ref{sec:h26a}. The best fit is achieved for the physical parameters shown in Table \ref{tab:inputparameters}. 

    \subsection{217 GHz and 342 GHz continuum maps} \label{sec:radiocontinuum}
    
        \begin{figure*}[ht!]
            \includegraphics[width=\textwidth]{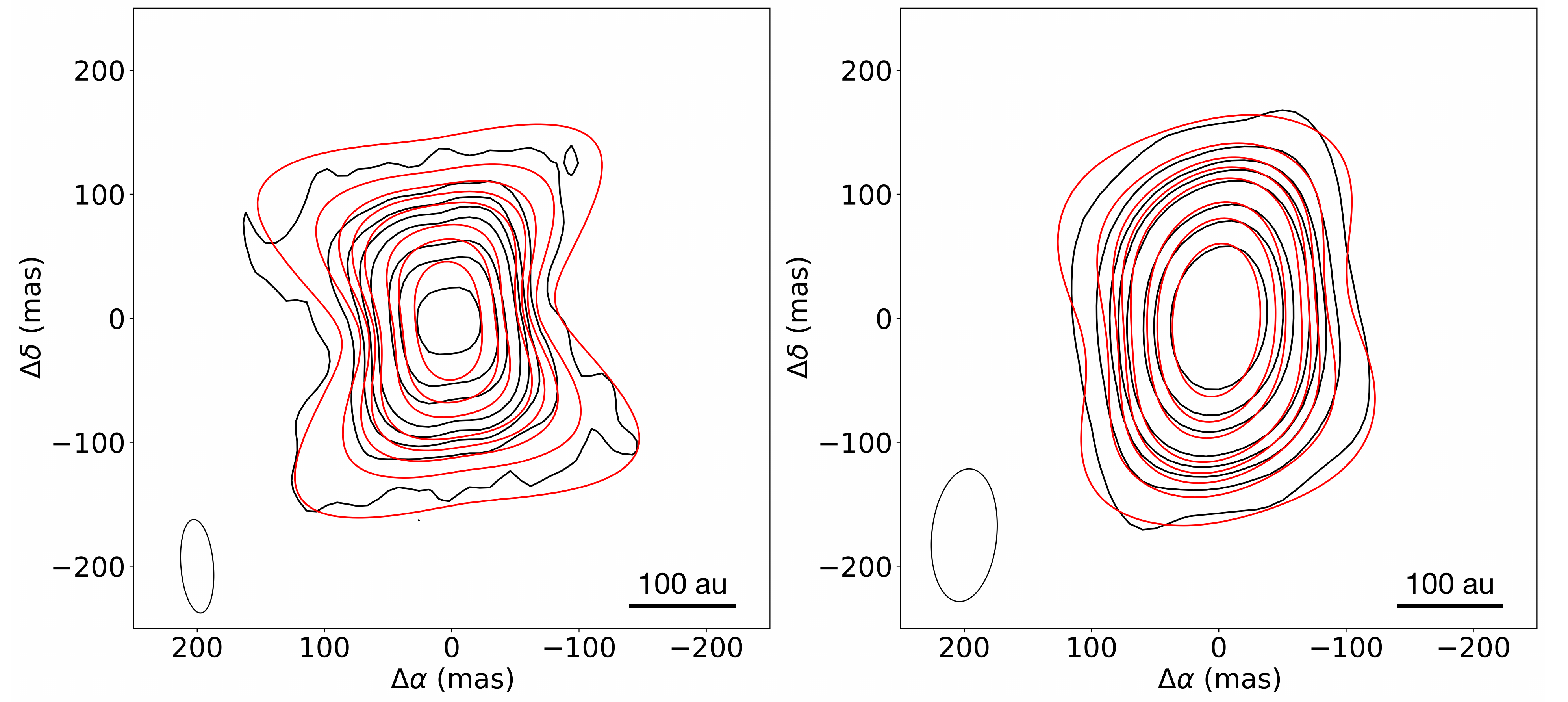}
            \centering
            \caption{Band 6 \mbox{217 GHz} and band 7 \mbox{342 GHz} continuum maps from ALMA (black contours) and model (red contours). The ellipse in the lower-left part of each image represents the synthesized beam of observations. Contour levels are [10, 30, 50, 70, 100, 200, 300, 500] times the RMS noise of the maps (0.48 mJy beam$^{-1}$ for band 6 and 1.39 mJy beam$^{-1}$ for band 7).
            \label{fig:continuum}}
        \end{figure*}
    
        \begin{table*}[]
            \caption{Total and peak flux of the continuum maps from ALMA bands 6 and 7, and from the modeled maps at the corresponding frequencies.}
            \begin{center}
            \begin{tabular}{|c|cc|cc|}
            \hline
            \multirow{2}{*}{Frequency (GHz)} & \multicolumn{2}{c|}{Integrated Flux (mJy)} & \multicolumn{2}{c|}{Peak Flux (mJy beam$^{-1}$)} \\ 
                                             & \multicolumn{1}{c}{ALMA}      & Model     & \multicolumn{1}{c}{ALMA}         & Model       \\ \hline
            217                            & \multicolumn{1}{c|}{1621}    & 1931    & \multicolumn{1}{c|}{388}        & 588       \\ \hline
            342                           & \multicolumn{1}{c|}{2755}    & 2628    & \multicolumn{1}{c|}{1853}        & 1866       \\ \hline
            \end{tabular}
            \end{center}
            \label{tab:radiocontinuum}
        \end{table*}
        
        \cite{prasad2023} have measured the flux and size finding an excellent agreement with the trends found from previous lower frequency observations \citep{tafoya2004}. In Fig. \ref{fig:continuum}, we present the contour maps of the modeled \mbox{217 GHz} and \mbox{342 GHz} continuum emission (red contours) superimposed on the observed maps (black contours). The modeled continuum maps have been obtained using the parameters of the best-fit model (see Table \ref{tab:inputparameters}) and smoothed to the same angular resolution of the ALMA observations for each frequency (75 $\times$ 25 mas, PA=4.4$^{\circ}$ for band 6 and 107 $\times$ 51 mas, \mbox{PA=-5.3$^{\circ}$} for band 7). As confirmed by \cite{prasad2023}, the \mbox{217 GHz} and \mbox{342 GHz} fluxes observed with ALMA are consistent with an spectral index close to $\alpha$ $\sim$0.6, which is also predicted in the model by \cite{baezrubio2013}. 

        Fig. \ref{fig:continuum} shows that the modeled continuum maps reproduce well the main features of the ALMA \mbox{217 GHz} and \mbox{342 GHz} continuum maps. The band 6 map shows the bipolar, hourglass shape also seen in the VLA maps of \citet{tafoya2004}, while in the band 7 continuum this shape is not resolved due to the larger size and south-north elongation of the synthesized beam, therefore showing the emission corresponding to a point source. The hourglass shaped nebula in the band 6 map is inclined $\sim$ 8$^{\circ}$ with respect to the east-west direction, also in agreement with previous observations. Note that the maps are not originally centered at the same coordinates due to the fact that the H26$\alpha$ data have been self-calibrated \citep{prasad2023}. We thus have used the continuum peak pixel of each observed and modeled map as the reference positions for Figure \ref{fig:continuum}. 
        
        The continuum flux is determined by the electronic density distribution $N_e(r,\theta)$ and electronic temperature of the ionized gas. For this model, we have considered the same electron density distribution and electron temperatures for the ionized disk and wind used by \cite{baezrubio2013} ($T_d$ and $T_0$, respectively; see Table \ref{tab:inputparameters}). As shown by \citet{baezrubio2013}, the predictions for the centroid map (Sect. \ref{sec:centroidmap}) are less sensitive to changes in the electron temperature than the predicted line profiles (Sects. \ref{sec:h30alineprofile}, \ref{sec:h26a}). An electron temperature for the wind of \mbox{$T_0$=12000 K} is required to explain the continuum flux and the intensities of the high-velocity humps observed for the H30$\alpha$ and H26$\alpha$ lines, while a temperature of \mbox{$T_d$=9450 K} is needed to best reproduce the intensities of the maser spikes arising from the disk. 
        
        In Table \ref{tab:radiocontinuum}, we compare the peak and integrated flux of the \mbox{217 GHz} and \mbox{342 GHz} continuum emission, with that measured by ALMA in bands 6 and 7. From \mbox{Table \ref{tab:radiocontinuum}}, we find that the values from the model are comparable to those from the observations. The integrated flux for band 6 in the model is slightly above that of the observations, which might be caused by the flux loss of emission that has been resolved out by the interferometer. For band 7, the integrated flux is strongly similar (with a deviation smaller than 5\%) between the model and the observations of the source, which in this case is not resolved.

    \subsection{Modeling the H30$\alpha$ centroid map} \label{sec:centroidmap}
    
\begin{figure}[h]
            \includegraphics[width=\columnwidth]{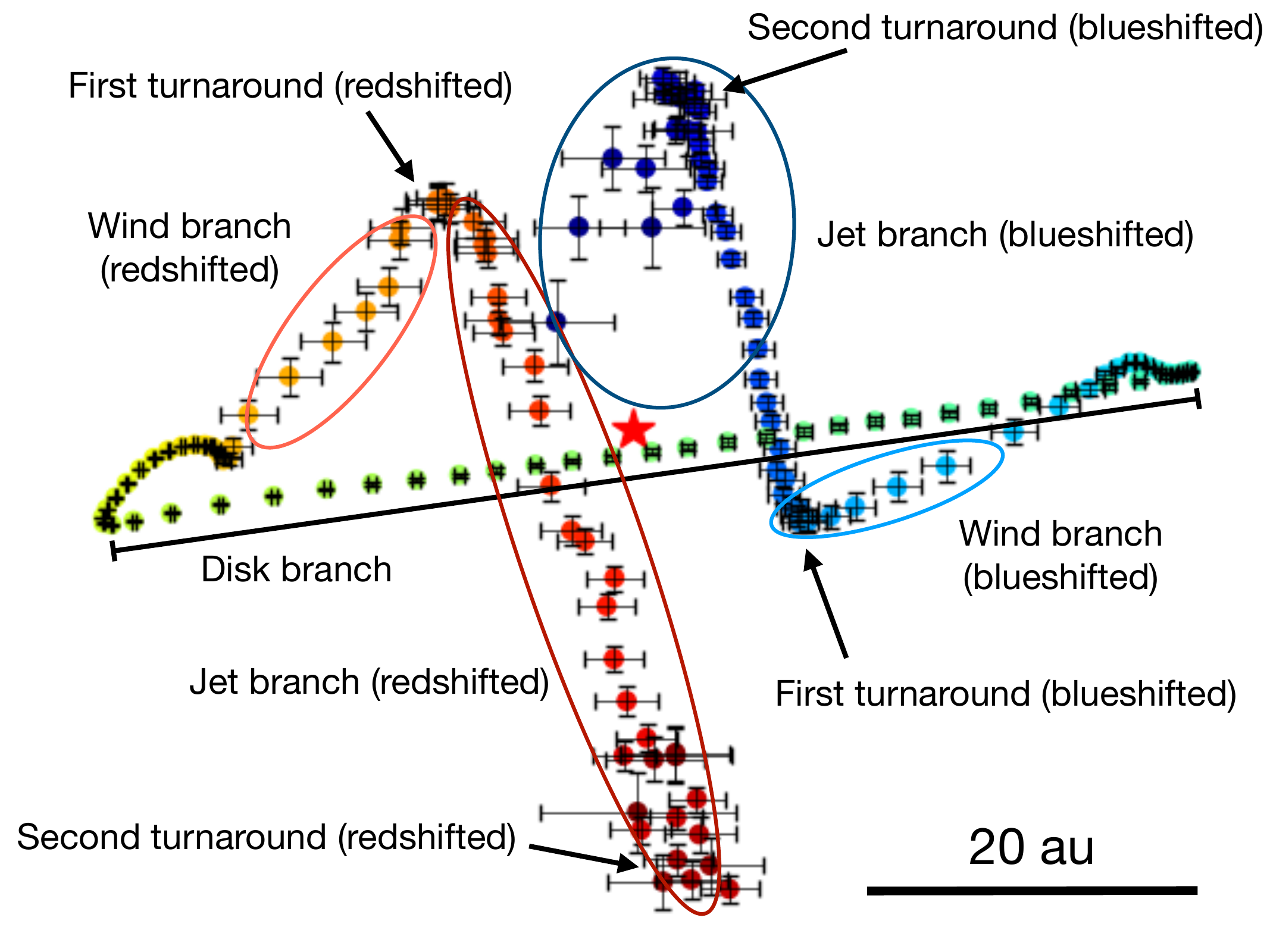}
            \centering
            \caption{Observed H30$\alpha$ centroids (colored points with errorbars) and an identification of the terms used in the discussion to describe the different parts of the map. Redder (bluer) colors represent more redshifted (blueshifted) emission. The centroid positions are located in the plane of the sky, where north is up and east is left: we have not shown the axes for clarity. See e.g. Fig. \ref{fig:panelcentroids} to see the axes. The red star symbol is located at the (0,0) position, i.e. the peak of the continuum emission.}
            \label{fig:branches}
    \end{figure}

     In the following subsections we describe how we have constrained the model by comparison with the ALMA H30$\alpha$ RRL centroids. Figure \ref{fig:branches} shows a sketch of the different parts that we identify in the centroid map for a definition of the terms that will be used along the text.
    
        \subsubsection{The ionized disk and wind in MWC 349A}
        
            \begin{figure}[ht!]
                \includegraphics[width=8.5cm, height=19.6cm]{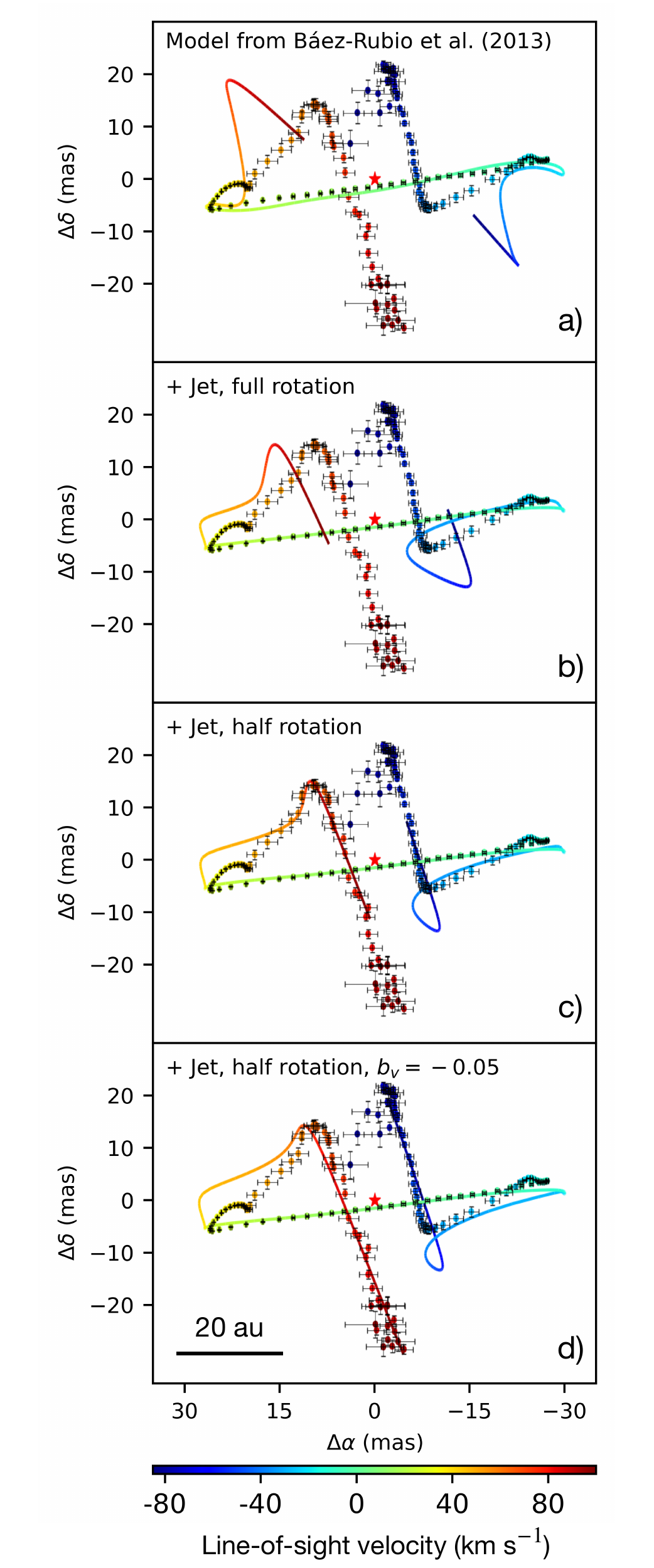}
                \centering
                \caption{Centroid maps of the H30$\alpha$ emission of MWC 349A from ALMA (dots with errorbars) and model (line). Panel a) shows the model in \cite{baezrubio2013}; in b) we incorporate a jet; in c), half of the outflow Keplerian rotation is removed; in d) the deceleration parameter $b_v=-0.05$ is included. The red star symbol is located at the (0,0) position, i.e. the peak of the continuum emission.
                \label{fig:panelcentroids}}
            \end{figure}
        
             In Figure \ref{fig:panelcentroids}, we report the comparison between the ALMA centroid map of the H30$\alpha$ RRL with the modeling results. We use the data from this RRL to constrain the physical parameters and the kinematics of the model because it does not suffer from saturation effects (see Section \ref{sec:h26a}). We present several cases to illustrate the effects of the different kinematic components considered in the MORELI code, including the new high-velocity jet and deceleration of the ionized wind (Sects. \ref{sec:jet}, \ref{sec:bv}).
                    
            In panel a) of Figure \ref{fig:panelcentroids}, we show a model based on the best fit obtained by \cite{baezrubio2013} using the PdBI data of \citet{martinpintado2011}, with minor changes in the input parameters (see Table \ref{tab:inputparameters}). This model includes a Keplerian rotating disk and an outflow expanding radially at a terminal velocity of 60 \unit{km.s^{-1}}
            in Keplerian rotation.
            
            The disk kinematic component in MWC 349A appears as points distributed in a straight line in the southeast-northwest direction in the centroid map, with radial velocities between -12 and 28 \unit{km.s^{-1}}. This set of centroids was defined as Group I in \cite{zhang2017}. These points have the lowest positional errors due to the strong maser amplification of the H30$\alpha$ emission, which is most efficient in the ionized disk where the optimum electronic density and temperature, and longest coherence lengths for amplification, are found \citep{planesas1992,baezrubio2013}. The position angle of this disk branch is $\sim$98$^{\circ}$, which is the same orientation as the disk as seen in radio continuum maps \citep{martinpintado1993,tafoya2004}. 
            
            The wind component in MWC 349A is probed by the centroids at velocities between 27 and 61 km s$^{-1}$ for the redshifted gas and between $-$35 and $-$12 km s$^{-1}$ for the blue-shifted emission. As shown in Figure \ref{fig:panelcentroids}, these centroids lift off the disk plane, until reaching a maximum value of declination $|\Delta\delta|$. These points, which are denominated Group II in \cite{zhang2017}, have been associated with the emission of the ionized wind launched from the surface of the disk \citep{martinpintado2011}; the model in \cite{baezrubio2013} could reproduce the H30$\alpha$ centroid positions of previous PdBI data by considering an ionized wind rotating in the same sense as the disk and expanding radially at a constant velocity. The highest radial velocity centroids obtained from previous observations in the PdBI \citep{martinpintado2011} and the SMA \citep{zhang2017} reached up to these velocities. We note, however, that the positional errors of previous PdBI and SMA data were up to six times larger than those of the ALMA dataset and therefore, the best-fit model of \cite{baezrubio2013} fails to reproduce the slope of the H30$\alpha$ gas that rises from the disk, see panel a) in Figure \ref{fig:panelcentroids}.
        
        \subsubsection{Discovery of a high-velocity jet in MWC 349A}\label{sec:introducingjet}
        
            In addition to the disk and wind components, the new ALMA data reveal an additional set of points, the jet branch, thanks to the exquisite sensitivity of the images. The additional H30$\alpha$ centroids of the jet branch appear at the highest radial velocities beyond 61 \unit{km.s^{-1}} and -35 \unit{km.s^{-1}} for the redshifted and the blueshifted line wing, respectively. This branch starts at the end of the wind branch, at the maximum values of declination $|\Delta\delta|$ where the centroids turn around towards the disk plane parallel to the rotation axis, eventually crossing the disk mid-plane (Figures \ref{fig:branches} and \ref{fig:panelcentroids}). There is a second turnaround back towards the disk along the polar axis in the opposite direction at the most extreme velocities (of $<-$80 \unit{km.s^{-1}} for the blueshifted line wing, and of $>$90 \unit{km.s^{-1}} for the redshifted emission).
        
            In panel b) of Figure \ref{fig:panelcentroids}, we present the new model where we have introduced a jet in the kinematics of the ionized gas as described in Sect. \ref{sec:jet}. From this Figure, it is clear that the model predictions are more similar to the ALMA observations in the sense that the modeled jet emission mimics the sense and amplitude of the downward (upward) excursions of the emission with high radial velocity shown in red (blue). The jet parameter with the strongest effect on the predicted centroid positions at high velocities is its orientation $(\theta_{jet},\varphi_{jet})$. In \mbox{Figure \ref{fig:jetorientations}} of the Appendix we show examples of modeled centroid maps with different jet orientations. We find that only a slightly tilted jet with the northern cone pointing towards the observer and the southern one expanding away can reproduce the observed jet branch. We discuss the possible physical origin of this tilt in Sect. \ref{sec:jettiltdiscussion}.
            
            The shift in the direction of the centroids leading to the start of the jet branch is asymmetrical: the redshifted loop is located at a higher value of $|\Delta\delta|$ than the blueshifted loop. This is explained by the larger amplification of the radiation in the northern cone due to the inclination $\theta_i$ of the disk towards the positive $z$ axis \citep{baezrubio2013}. However, we need to introduce an asymmetry between the orientation of the northern and the southern jet to accurately describe the second turnaround (see Figure \ref{fig:asymmetricaljet} in the Appendix). The northern cone has an angular deviation of 16$^{\circ}$ with respect to the disk rotation axis, while the southern cone shows a deviation of 22$^{\circ}$ (see Table \ref{tab:inputparameters}). A sketch of the jet geometry in the final model is shown in the leftmost panel of Figure \ref{fig:jetorientations} in the Appendix. 
                    
            The velocity $v_{jet}$ and its collimation $\psi_0$ are also key to model the H30$\alpha$ jet centroids since they have a strong influence on the prediction of the radial velocity range revealed by the observations. Our best-fit model in \mbox{Figure \ref{fig:panelcentroids}} has a jet velocity of 250 \unit{km.s^{-1}} with \mbox{$\psi_0$ = 0.4} (Table \ref{tab:inputparameters}). This implies that the velocity of the jet reaches half its maximum velocity at an angular distance from the jet axis of 24$^{\circ}$, which gives a rather poorly collimated jet when compared to others, e.g. the one in \mbox{Cep A HW2} \citep[semi-opening angle of \mbox{$\sim9^{\circ}$},][]{torrelles2011}. Models with jet velocities under this value do not fit satisfactorily the observed data, regardless the collimation. However, we note that faster jets with narrower collimations also provide good fits. For instance, we find that a model including a jet expanding at \mbox{540 \unit{km.s^{-1}}} with a collimation $\psi_0=0.2$ and with the same orientation as described before, provides good results. This would be consistent with the jet velocity of \mbox{$\sim$ 575 \unit{km.s^{-1}}} found by \cite{prasad2023} using the highest radial velocities sampled by ALMA.
            
            Although the jet produces H30$\alpha$ maser features that match closely the ALMA observations, no combination of orientation and velocity is able to reproduce satisfactorily the location of the jet branches. To bring the branches closer to the rotation axis of the disk, we need to change the rotation of the wind component to half of the disk rotation in the model, as shown in panel c) of Figure \ref{fig:panelcentroids}. For an example of the behavior of the modeled centroid map with the outflow rotation following that of the disk see Figure \ref{fig:centroidsrot} in the Appendix. The change in the wind rotation is discussed in Sect. \ref{sec:rotationdiscussion}.              
        
        \subsubsection{Deceleration of the ionized wind expansion}
        
            The result of the incorporation of the deceleration parameter $b_v$ (Sect. \ref{sec:bv}) in the outflow expansion is presented in panel d) of Figure \ref{fig:panelcentroids}. As in the case of the collimation, different combinations of the jet velocity $v_{jet}$ and the radial deceleration $b_v$ yield similar results. We constrain these values by adjusting the length of the jet branch until it reaches the second turnaround at the most extreme velocities, where the centroids turn back to the disk mid-plane in the direction of the rotation axis. Before including the deceleration, the high velocity emission located further in the jet branch does not appear in the centroid map since its velocity is outside the velocity range of the observations. With the incorporation of $b_v$, the jet now moves at a slower velocity, bringing its velocity inside the observed range resulting in the extension of the length of the branch.
                    
            The shape of the wind branch is very sensitive to small changes in $b_v$: higher values (in magnitude) make the wind branch less steep for higher velocities. This is clearly illustrated in panels c) and d) of Figure \ref{fig:panelcentroids}: in panel c), where $b_v = 0$, the wind branch is accurately reproduced, especially in the redshifted part. Then in panel d), we have set $b_v = -0.05$ (Table \ref{tab:inputparameters}), a very small value that is enough to extend the length of the jet branch to match the observations. However, the wind branch is less steep, with a somewhat more rounded profile than in panel c), and this effect grows with larger values of $b_v$. We have found a compromise between reproducing the shape of the wind branch and the extent of the jet branch. Had we increased the magnitude of $b_v$ even more, we would have reproduced the second turnaround towards the disk mid-plane at the most extreme velocities, as shown by the model for an extended range of radial velocities (see Figure \ref{fig:centroidsfullrange} in the Appendix); but that would have had the cost of having a strong mismatch in the wind branch.
                
    \subsection{Constraining the model with the H30$\alpha$ line profile}\label{sec:h30alineprofile}
            
        \begin{figure*}
            \includegraphics[width=18cm, height=12cm]{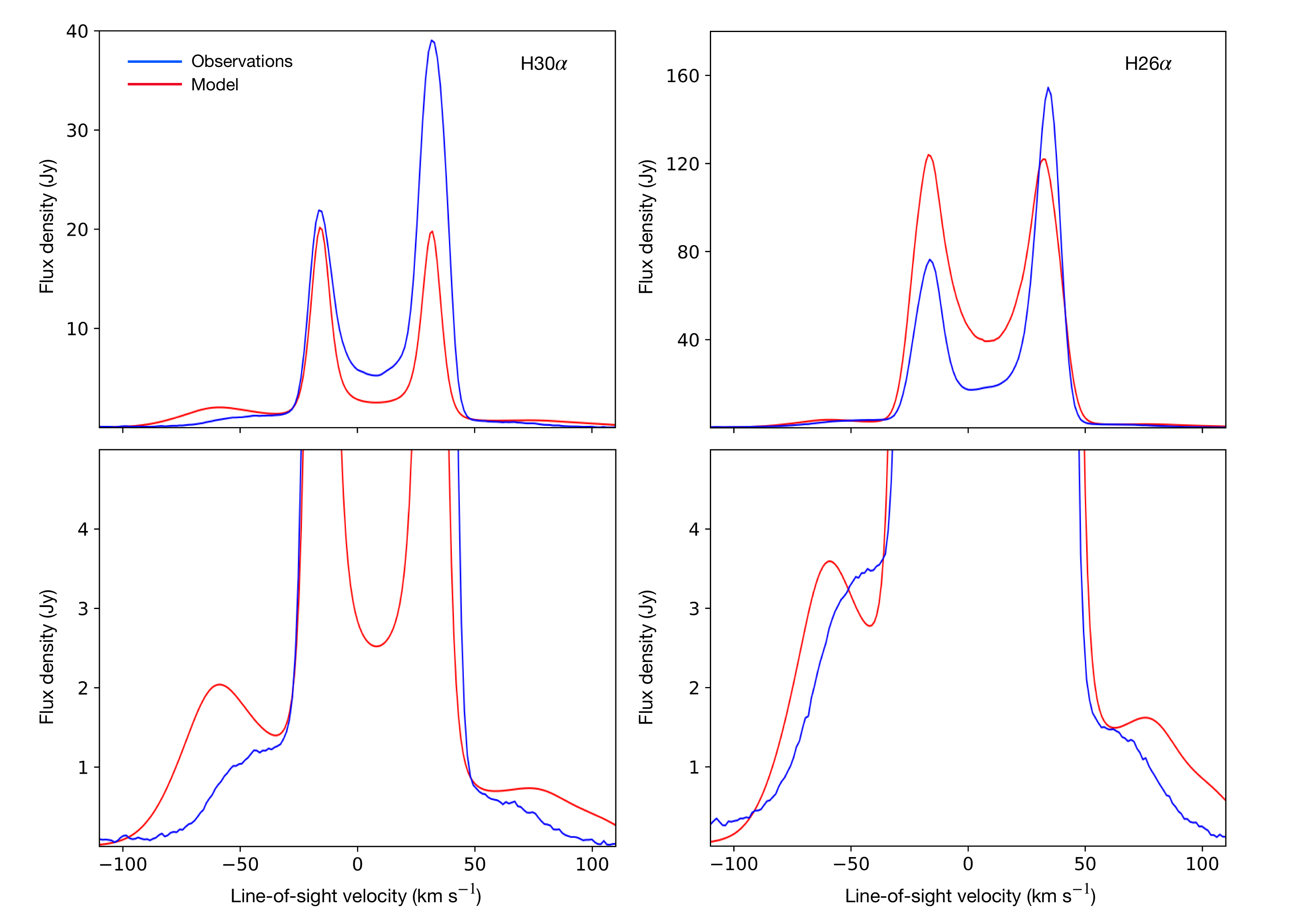}
            \centering
            \caption{Line profile of the H30$\alpha$ (left pannels) and H26$\alpha$ (right pannels) from observations (blue) and the final model (red). The lower panel is a zoom in at the lower intensity values from the upper panel. The double peak traces emission from the ionized disk, while the weaker features at higher radial velocities (the "wing humps") arise from the outflow. The minimum intensity of the double peak is located at the systemic velocity of the source, 8 \unit{km.s^{-1}}.  \label{fig:lineprofile}}
        \end{figure*}
        
        Simultaneously to the analysis of the centroid maps, the model also predicts the spatially integrated H30$\alpha$ line profile and we compare it with the line profile observed toward MWC 349A (left panels of Figure \ref{fig:lineprofile}), which is obtained from the amplitude of the centroid fit \citep{prasad2023}. We have used the LTE departure coefficients $b_n$ from \cite{storeyhummer1995} in our modeling as in \cite{baezrubio2013}.
        
        The H30$\alpha$ emission consists of a double-peaked line profile with the maxima centered at velocities of \mbox{$\sim$ -15} and \mbox{$\sim$ 32 \unit{km.s^{-1}}} for the blueshifted and redshifted components, respectively. The H30$\alpha$ line profile also shows broad, weaker shoulders at higher velocities (between \mbox{-30 and -70 \unit{km.s^{-1}}} in the blueshifted part and between \mbox{50 and 80 \unit{km.s^{-1}}} for the redshifted part) which we refer to as ``wing humps''.
        
        The double-peaked H30$\alpha$ emission arises from the ionized, Keplerian rotating disk; the peak intensities correspond to the maser spikes in the disk, where the conditions for the maser amplification are optimum together with largest coherent length \citep{planesas1992}. The peak-to-peak distance is mainly affected by the central mass of the star $M_*$ and by the radius of the ionized, Keplerian rotating disk $r_K$. A higher central mass widens the peak-to-peak separation, as expected for Keplerian rotation; larger radii of the Keplerian ionized disk result in a smaller separation of the peaks. This radius has a strong effect on the centroid map: a large value results in a more extended disk branch. Therefore, these parameters have been constrained to $M$ = 23 M$_\odot$ and to $r_k$ = 46 au, by finding a satisfactory fit to both the H30$\alpha$ line profile and the centroid map. The values used for the final model can also be found in Table \ref{tab:inputparameters}.
        
        As for the wing humps, they are mostly affected by the parameters of the wind and the jet. The original model without jet does not fully reproduce the velocity width of the wing humps, as they are not broad enough, see Fig. \ref{fig:h30alinewithoutjet} in the Appendix. With the addition of the jet, the wing humps broaden, in agreement with the observed line profile. The jet peak velocity $v_{jet}$ and orientation $(\theta_{jet},\varphi_{jet})$ change the location of the wing humps, i.e. faster jets or those with a higher projection onto the line of sight result in wing humps located at larger radial velocities. Therefore, our best fit (by eye) to the centroid maps and line profiles gives a jet with a velocity of \mbox{250 \unit{km.s^{-1}}} and a collimation parameter of $\psi_0=0.4$. As mentioned in Sect. \ref{sec:introducingjet}, the best fit is found with an asymmetrical jet, with the northern lobe tilted 16$^{\circ}$ from the rotation axis, and a tilt angle of 22$^{\circ}$ for the southern lobe (Table \ref{tab:inputparameters}).

        The intensity of the double peak and the wing humps depends on the opening angle of the ionized disk $\theta_d$, while the electron temperature of the disk $T_d$ only affects the double peak emission. Lower values of $\theta_d$ increase the height of the wing humps and lower the intensity of the double peak line. Higher $T_d$ decreases even further the double peak emission. The best-fit values of our model are listed in Table \ref{tab:inputparameters}. We note that the ionized disk and the ionized wind can have different electron temperatures due to the different cooling produced by higher electron density in the disk \citep[][Sect. \ref{sec:model}]{baezrubio2013}. In this model the disk is considered to have a lower electron temperature than the ionized wind, as constrained from the radiocontinuum measurements (Sect. \ref{sec:radiocontinuum}).

        Our model also allows to check whether the H30$\alpha$ line is consistently amplified at all velocities due to non-LTE effects. Stimulated emission takes place when the total optical depth from the line $\tau_{\nu,l}$ and the continuum $\tau_{\nu,c}$ lies between \mbox{-1} and 0, i.e. \mbox{$ -1 < \tau_{\nu,l} + \tau_{\nu,c} < 0$}. On the other hand, the emission is considered to experience maser amplification when \mbox{$\tau_{\nu,l} + \tau_{\nu,c} \ll -1$} \citep{baezrubio2013}. We find that up to 1\% of all the lines of sight considered in the model suffer from maser amplification for the velocity ranges \mbox{(-68,-44) \unit{km.s^{-1}}}, i.e. the blueshifted wing hump; \mbox{(-20, 30) \unit{km.s^{-1}}}, corresponding to the double peak from the ionized disk emission; and \mbox{(64,80) \unit{km.s^{-1}}} for the redshifted wing hump, less intense and therefore with a narrower range than the blueshifted one. This low fraction of maser-amplified lines of sight dominates the line emission, as seen in the line profile. Outside these velocity ranges, the line is consistently dominated by stimulated emission.

    \subsection{Modeling the H26$\alpha$ centroid map and line profile: saturation effects}\label{sec:h26a}
    
        \begin{figure}[ht!]
                \includegraphics[width=\columnwidth]{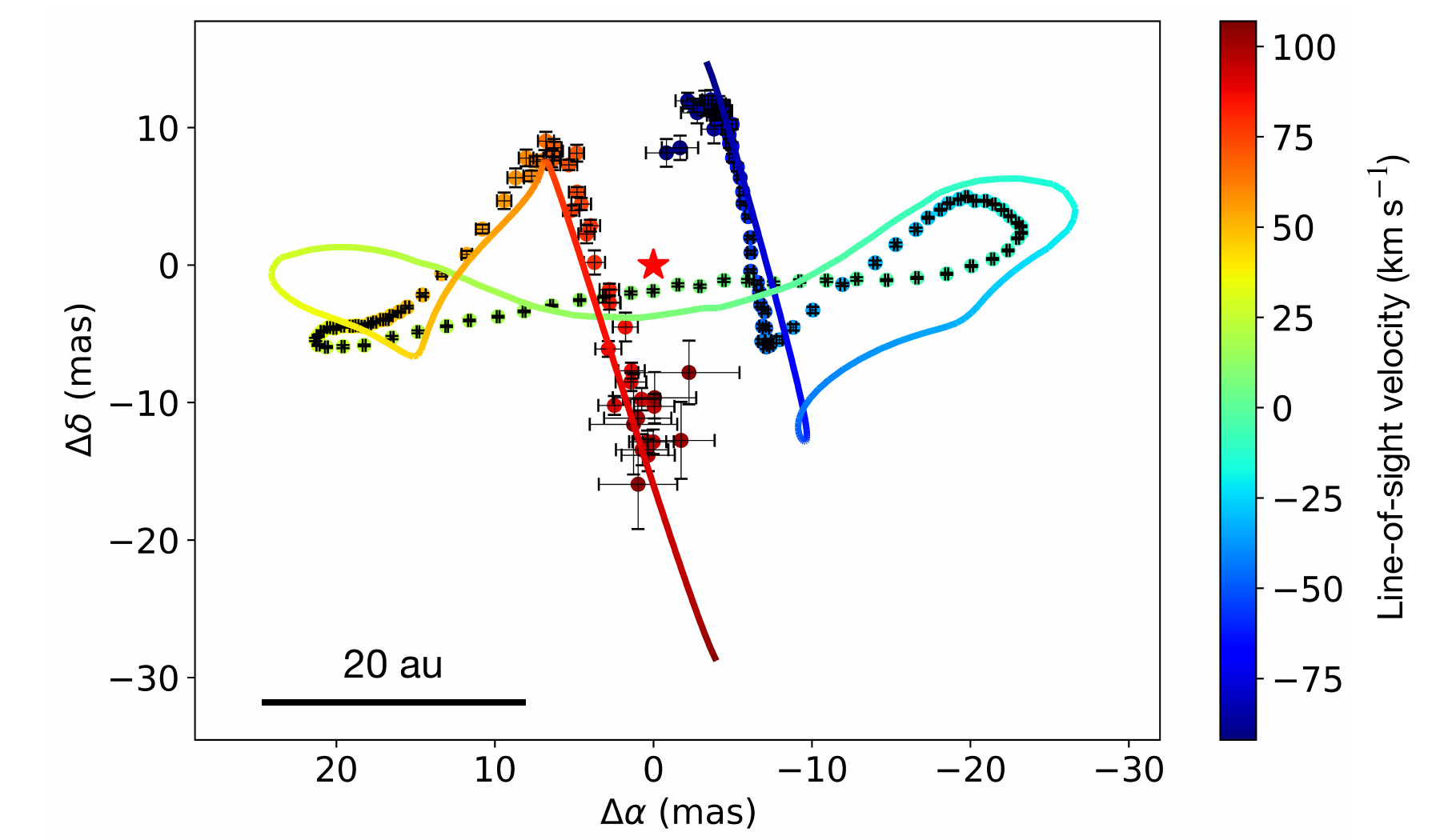}
                \centering
                \caption{Centroid map of the H26$\alpha$ emission of MWC 349A from ALMA (dots with errorbars) and the final model (line). The red star symbol is located at the (0,0) position.
                \label{fig:h26acentroids}}
        \end{figure}
        
        We consider the best fit parameters derived from the H30$\alpha$ observations presented in Sects. \ref{sec:centroidmap} and \ref{sec:h30alineprofile} and shown in Table \ref{tab:inputparameters} to predict both the centroid map and the line profiles of the H26$\alpha$ emission. The right panels of Figure \ref{fig:lineprofile} show the modeled and the measured integrated line profiles of the H26$\alpha$ line emission, and the comparison of the predicted and the observed centroid maps is shown in Fig. \ref{fig:h26acentroids}. Following the same analysis we have performed for the H30$\alpha$ line (Sect. \ref{sec:h30alineprofile}), we find that the H26$\alpha$ line emission is maser-amplified from \mbox{-80} to 100 \unit{km.s^{-1}}, with stimulated emission taking place outside this range.
        
        The H26$\alpha$ maser emission, unlike H30$\alpha$, is strongly saturated in the disk branch \citep{baezrubio2013,tran2021}, while the wind and jet branches do not suffer from saturation since the maser amplification is still moderate. Our radiative transfer code considers saturation along each line of sight \citep{baezrubio2013}. Two parameters in the model are used to describe saturation: the solid angle of the maser beam, which we have adopted to be $4\pi/\Omega$ = 60 \citep{thum1994}, and the degree of saturation above which the regime turns to be linear instead of exponential, $J_{\nu}/J_{\nu, sat}$ = 15, which we have chosen roughly by adjusting the modeled peak intensity of the H26$\alpha$ line profile to the observed one. A more detailed explanation on the implementation of saturation in the MORELI code is given in \cite{baezrubio2013}.
        
        Regarding the centroid map in Fig. \ref{fig:h26acentroids}, the three branches defined in Sect. \ref{sec:centroidmap} are present as for the H30$\alpha$ model, although there are some mismatches with observations. We discuss the limitations of the H26$\alpha$ modeling in Sect. \ref{sec:h26adiscussion}.
        
    \subsection{Rotation curves of the H30$\alpha$ and H26$\alpha$ emission}\label{sec:rotationcurves}

        \begin{figure}
            \centering
            \includegraphics[width=8.5cm, height=14.5cm]{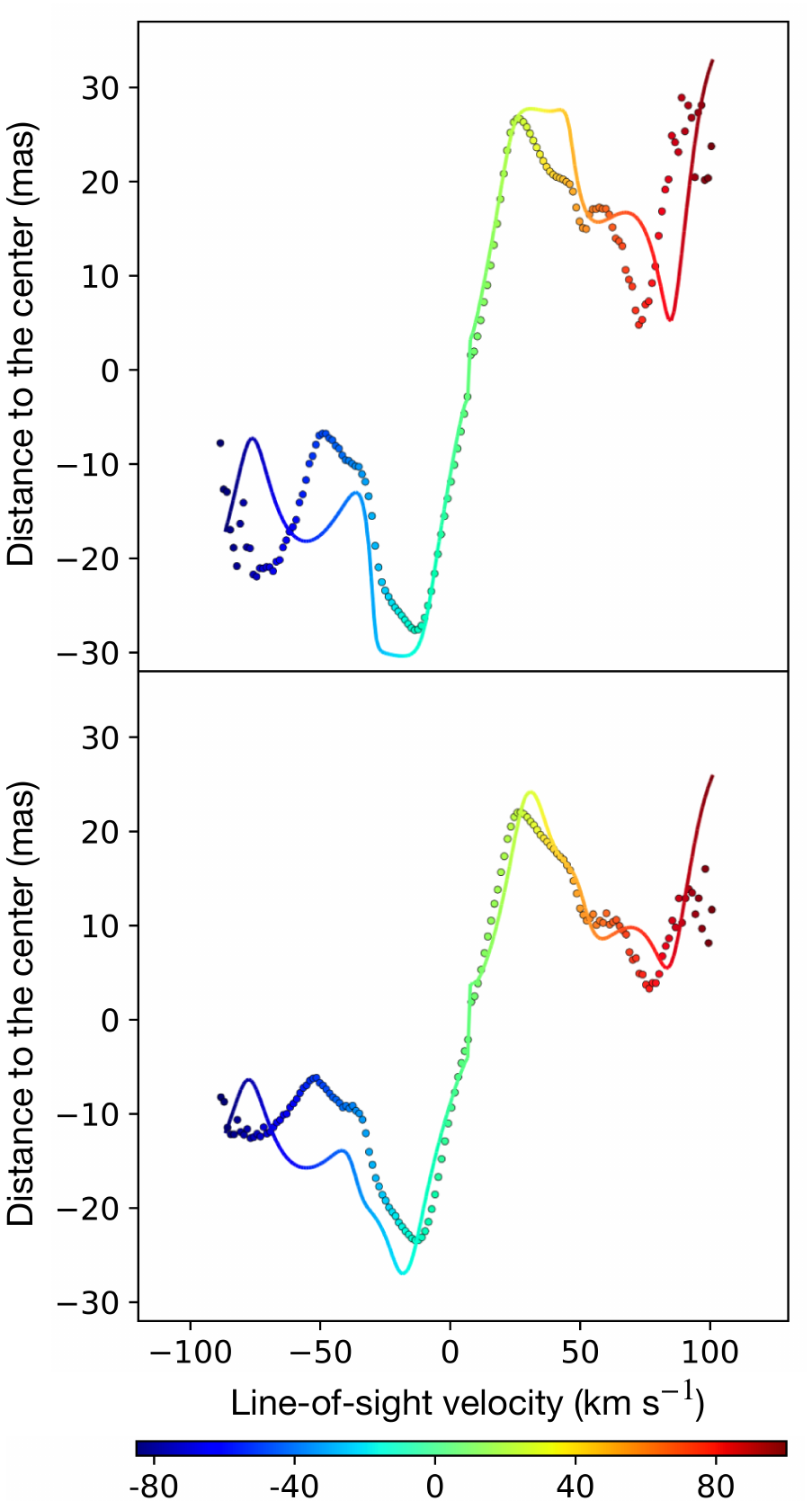}
            \caption{Rotation curves of the H30$\alpha$ (top) and H26$\alpha$ (bottom) centroids. The distance from the center is calculated as $\sqrt{\Delta\alpha^2 + \Delta\delta^2}$. Observations are represented with circles, while the model results is represented by the colored line.}
            \label{fig:pv}
        \end{figure}

        The predicted and observed rotation curves of the H30$\alpha$ and H26$\alpha$ lines are presented in Fig. \ref{fig:pv}. This is another way of representing the information contained in the centroid maps, where here the match of the radial velocities between the model and the kinematics can be analyzed in more detail.
        
        From Fig. \ref{fig:pv}, we find that the Keplerian rotation of the disk is seen as a straight line at velocities between -12 and 28 \unit{km.s^{-1}}. An estimation of the central mass can be done with the slope of this line \citep{zhang2017,prasad2023}. However, this underestimates the actual mass of MWC 349A because the size of the maser disk is underestimated by the centroid fit, since it corresponds to the averaged position of the extended emission. As discussed in Sect. \ref{sec:h30alineprofile}, we have constrained the mass of the MWC 349A source to 23 M$_{\odot}$ (Table \ref{tab:inputparameters}) by using the peak-to-peak distance measured in the unsaturated H30$\alpha$ and the H26$\alpha$ RRL profile.
        
        Despite the general agreement between the modeled and observed data points in Fig. \ref{fig:pv}, a clear shift is seen at high velocities for both lines, especially for the blueshifted gas. We discuss the possible reasons behind this misalignment in Sects. \ref{sec:jetdiscussion} and \ref{sec:bvdiscussion}.

    \subsection{Mass loss and momentum extraction by the jet and the wind} \label{sec:massmomentum}

        Once the jet and the wind expansion is constrained, we can extract from the model the information about their mass, mass loss rate, and linear and angular momentum. 

        To do this, we have first computed the mass of the ionized disk, wind and jet components. This is calculated by adding the mass inside all cells of size \mbox{$dx \times dy \times dz$} for every component. In the case of the ionized disk, we obtain its mass by considering the cells inside the ionized layer between the angles \mbox{$(\theta_a-\theta_d, \theta_a)$} (Table \ref{tab:inputparameters}), yielding a value of \mbox{$\sim2\times10^{-4}\,$M$_{\odot}$}. The mass of the ionized wind is calculated by considering the cells within the wind double cone, with a value of \mbox{$M_{wind}\sim1\times10^{-4}$ M$_{\odot}$}. Similarly, we compute the ionized mass in the jet by considering the cells with a jet velocity component greater than half the maximum velocity $v_{jet}$, i.e. with a semi-opening angle lower than 24$^{\circ}$ (see Equation \ref{eq:psiweight}), obtaining \mbox{$M_{jet}\sim2\times10^{-5}$ M$_{\odot}$}. This value is a factor of 5 lower than the ionized wind mass. This is however expected given the orientation of the jet close to the rotation axis of the disk ($\sim$20$^{\circ}$), and the density distribution in the double cone, which entails a higher density near the walls of the cone (Sect. \ref{sec:model}).
        
        Using the same approach as employed in \cite{baezrubio2013}, MORELI integrates up to the effective radius that contains the free-free emission of a partially optically thick wind as derived by \cite{panagiafelli1975}. The corresponding semi-cone length is \mbox{172 au}, which implies dynamical timescales for the jet and the wind of $172\text{ au}/v=$ 3.3 and 13.7 years, respectively. Taking this into account, we estimate a mass loss rate for the jet of \mbox{$\dot{M}_{jet}\sim6\times10^{-6}$ M$_{\odot}$ yr$^{-1}$}. This value is comparable to the one found for the jet powered by the massive protostar \mbox{Cep A HW2}, which ranges between \mbox{$\dot{M}=1.5\times10^{-6}$ M$_{\odot}$ yr$^{-1}$} \citep{carrascogonzalez2021} and \mbox{$\dot{M}=3.2\times10^{-6}$ M$_{\odot}$ yr$^{-1}$} \citep{jimenezserra2011}. For the wind, the mass loss rate can be computed from the continuum flux (Table \ref{tab:radiocontinuum}) and using Equation 19 in \cite{reynolds1986}, yielding a value of \mbox{$\dot{M}_{wind}=5\times10^{-5}$ M$_{\odot}$ yr$^{-1}$}. This large difference in mass loss rate, and hence in continuum flux density, might explain why the jet cannot be identified in the radiocontinuum images of Figure \ref{fig:continuum} \citep{prasad2023} or in the ones from \cite{tafoya2004}.

        To compute the wind and jet momentum, we apply a similar method obtaining values of \mbox{$(Mv_r)_{wind}\sim5\times10^{-3}$ M$_{\odot}$ \unit{km.s^{-1}}} and \mbox{$(Mv_r)_{jet}\sim3\times10^{-3}$ M$_{\odot}$ \unit{km.s^{-1}}}, respectively. The derived momentum rates are \mbox{$(\dot{M}v_r)_{wind}\sim4\times10^{-4}$ M$_{\odot}$ yr$^{-1}$ \unit{km.s^{-1}}} and \mbox{$(\dot{M}v_r)_{jet}\sim1\times10^{-3}$ M$_{\odot}$ yr$^{-1}$ \unit{km.s^{-1}}}, respectively. The jet and the wind drive similar amounts of linear momentum. For comparison, the momentum rate of the \mbox{Cep A HW2} jet is \mbox{$\sim 3 \times 10^{-4}$ M$_{\odot}$ yr$^{-1}$ \unit{km.s^{-1}}}  \citep{curiel2006}, very similar to the one reported here for MWC 349A.

        Finally, the derived angular momentum of the disk, wind and jet are, respectively, \mbox{$(Mv_{\varphi}\rho)_{disk}\sim3\times10^{-2}$ M$_{\odot}$ \unit{km^{2}.s^{-1}}}, \mbox{$(Mv_{\varphi}\rho)_{wind}\sim6\times10^{-3}$ M$_{\odot}$ \unit{km^{2}.s^{-1}}} and $(Mv_{\varphi}\rho)_{jet}\sim1\times10^{-3}$ M$_{\odot}$ \unit{km^{2}.s^{-1}}, where $\rho$ is the radial distance of the cell to the rotation axis of the disk. These values show that the wind has extracted $\sim$6 times more angular momentum from the system than the jet is extracting, and that once combined, they have extracted about 25\% of the angular momentum of the disk.

\section{Discussion} \label{sec:discussion}

    \subsection{Origin of the jet embedded within the ionized wind} \label{sec:jetdiscussion}
    
        Our modeling results have shown that in order to reproduce the H30$\alpha$ and H26$\alpha$ centroid maps, we need to introduce a new kinematic component for the ionized gas in MWC 349A: a jet that fills the double cone of the ionized wind and that expands and rotates together with the wind. As already constrained by \citet{baezrubio2014}, the wind is launched from the disk at ~24 AU from the star with an expanding velocity of 60 \unit{km.s^{-1}}. This wind has been explained by a magneto-hydrodynamical launching mechanism \citep{martinpintado2011,baezrubio2014,zhang2017} and it is compatible with the disk wind scenario of \cite{koniglpudritz2000}. Our finding of a faster \mbox{(250 \unit{km.s^{-1}})} component rotating in the same sense as the disk and which is collimated near the rotation axis is also compatible with such models. We have checked that this jet is consistent with the mass loss and momentum rates of another well-known massive system powering a jet, \mbox{Cep A HW2} (Sect. \ref{sec:massmomentum})
        
        Accretion of material in young stars has been long observed to be connected to ejection \citep[e.g.][]{bally2016}, which extracts excess angular momentum from the infalling matter and allows accretion toward the star. The evolutionary status of MWC 349A is under debate: it could be a massive young star still accreting matter \citep[e.g.][]{cohen1985,strelnitski2013}, or an evolved star (e.g. \citealt{kraus2020} and references therein). \cite{zhang2017} showed that the ionized wind removes angular momentum from the disk, which leads to accretion onto the central star; the finding of a jet is in agreement with this result. If MWC 349A is a young star, it would still be actively accreting mass from its disk, despite the photoionization of its neutral circumstellar material. 
        
        Accretion is not necessarily steady: it can be irregular or episodic, which would result in episodic ejections as well. Our jet model does not account explicitly for such phenomenon, but it is possible that could be behind the variability of the RRL emission \citep{martinpintado1989,thum1992} and of some of the mismatches between our model and observations. For instance, the model fails to reproduce exactly the highest velocities in the RRL emission. In the centroid map, the final model (panel d) in Fig. \ref{fig:panelcentroids} ) does not completely match the extreme velocity centroid positions: the second turnaround towards the disk mid-plane at the highest velocities is not seen in the model unless we extend the radial velocity range (Fig. \ref{fig:centroidsfullrange} at Appendix), which means that the jet in the model is expanding faster than the actual jet at these locations near the rotation axis. This is also evident in the line profile (Fig. \ref{fig:lineprofile}), where the wing humps of the model are slightly shifted towards larger velocities than the observed ones, and in the rotation curves (Fig. \ref{fig:pv}), where there is a clear offset between the model and the observations at high velocities, especially in the blueshifted part. The ionized material could have been ejected in different episodes with slightly different speeds, which would not be accounted for in our jet described by a single expansion velocity peak value and a single-axis cone. 
        A sudden ejection \citep[as recently found in e.g. \mbox{S255 NIRS 3},][] {carattiogaratti2017, cesaroni2018} could thus be the origin of the high-velocity jet. 
        
        Finally, note that disk winds are launched over a wide range of radii \citep{blandfordpayne1982}, in contrast to X-winds \citep{shu1994}, where the jet is launched from the inner edge of the accretion disk. This 'onion-like' structure, if present on the actual jet, could have its signature in our model as the collimation that implies lower expansion velocity values for the jet at larger angular distances from the jet orientation (Sect. \ref{sec:jet}), which is observed in other jets \citep{frank2014}.

    \subsection{Tilted jet} \label{sec:jettiltdiscussion}

        From Section$\,$\ref{sec:results}, we find that MWC 349A's jet appears deviated $\sim 20^{\circ}$ with respect to the symmetry axis of the disk, with a tilt degree difference of $6^{\circ}$ between the northern and the southern lobes (Table \ref{tab:inputparameters}). Asymmetrical jets have also been found toward other objects such as the massive protostar \mbox{Cep A HW2}, where the northern and southern parts of the jet are not aligned \citep{carrascogonzalez2021}; and the Herbig Ae/Be star RCrA, composed of a triple system and a dust cavity perpendicular to the circumstellar disk where a high-velocity jet and possibly a disk wind are present \citep{rigliaco2019}.
        
        Asymmetrical jets can be produced by different mechanisms: turbulent accretion during the first stages of its formation \citep[e.g.][] {rosenkrumholz2020}; precession \citep[e.g.][]{bally2007,fendtzinnecker1998} or warping of the disk due to the interaction with companions \citep[e.g.][]{alylodato2020};
        or the interaction with the surrounding medium and/or with an external magnetic field \citep{fendtzinnecker1998}. Magneto-hydrodynamical (MHD) simulations of the formation process of a protostar predict random orientations of the outflow if there is a misalignment between the rotation axis of the system and the global magnetic field direction \citep{machida2020}. However, the magnetic field in MWC 349A is found to be parallel to the disk rotation axis from mid-infrared spectropolarimetry \citep{aitken1990}, which makes this possibility less plausible.

        Alternatively, in the multiple system scenario, the interaction with a stellar companion can trigger the precession of the disk \citep[e.g.][]{bally2007} and generate warps \citep[e.g.][]{alylodato2020}, both of which can deviate the direction of a jet launched from the disk. It has been proposed that MWC 349A may be part of a binary system with the nearby star MWC349B at a projected separation of 2.4" \citep{brugelwallerstein1979}. \citet{cohen1985} and \citet{tafoya2004} suggested that the morphology of the continuum emission of MWC 349A at 2 cm (which forms an arc-like shape) was evidence of the interaction between these two sources, supporting the idea that they are physically bound. However, \cite{meyer2002} found that the star MWC349B is more polarized than MWC 349A, indicating that the former is located far behind the latter, and therefore their close distance in the plane of the sky may be due to a projection effect. More recently, \cite{drew2017} performed high resolution optical spectroscopy of MWC349B concluding that the two stars are not bound as inferred from the large difference of their radial velocities. 
        
        We note that the star MWC 349A has been suggested to be a close unresolved binary inside a circumbinary disk \citep{jorgenson2000,hofmann2002,gvaramadzementen2012}. Direct evidence for a close companion is, however, still lacking. The presence of a companion could induce 
        the warping of the MWC 349A disk, as recently proposed for the massive protostar \mbox{Mon R2-IRS2} \citep{jimenezserra2020}. Indeed, the H21$\alpha$ centroid positions of the disk branch in \mbox{Mon R2-IRS2} clearly deviated from the disk plane. In the case of MWC 349A, the H30$\alpha$ disk emission is distributed in an almost perfect straight line, but the H26$\alpha$ disk centroids deviate slightly from the disk plane, especially in the blueshifted part (see Fig. \ref{fig:h26acentroids}). It is thus unclear whether the kinematics of the inner regions of the disk probed by the H26$\alpha$ line are affected by the presence of a companion or whether the warping of the centroids in the disk branch is simply caused by saturation of the H26$\alpha$ line. We note that the warping predicted by the current model presents an opposite behaviour to the one of the observations.
        
    \subsection{Jet opening and collimation} \label{sec:jetopeningdiscussion}
    
        In our model, the jet is embedded within the ionized wind, i.e. within the double cone of semiaperture $\theta_a-\theta_d = 44^{\circ}$ (see Table \ref{tab:inputparameters}). The collimation has a value $\psi_0 = 0.4$ that implies that the jet velocity has half the peak value at 24$^{\circ}$ (calculated by substituting these values in Equation  (\ref{eq:psiweight})), which yields a poorly-collimated jet. For massive stars, more evolved systems could hold less collimated jets as they grow in mass and stellar radiation as proposed by \cite{beuthershepherd2005} and \cite{arce2007}, although some counter-examples have been found too \citep{li2014}. The evolutionary sequence proposed by \cite{beuthershepherd2005} suggests that evolved, massive stars in the UCHII phase would host remnants of the initial, very collimated jet that was present at the early stages of the formation. The remnant jet would maintain a small degree of collimation while being coupled with a lower velocity wind. This evolutionary sequence agrees with the correlation between the age of massive young stellar objects (MYSOs) and the opening angles of bipolar outflow cavities found by \cite{frost2021} using near infrared observations, where older sources present wider outflow cavities. 
        
        As stated above in Sect. \ref{sec:jetdiscussion}, the evolutionary status of MWC 349A is under debate, despite the evidence of ejection of material from the circumstellar disk. If the bipolar outflow of MWC 349A was carved out into the envelope material by a jet in a previous evolutionary phase, the finding of a poorly collimated jet embedded within the wind would be compatible with the scenario of a young star in an advanced phase of its formation. 
        
    \subsection{Rotation of the outflow} \label{sec:rotationdiscussion}
    
        The rotation of the wind and jet has been decreased to half of the disk rotation in our final model to bring the jet branches closer to the rotation axis of the system in the centroid map (Sect. \ref{sec:centroidmap}; Fig. \ref{fig:panelcentroids} and Fig.  \ref{fig:centroidsrot} in the Appendix). The consideration of the Keplerian rotation in the kinematics of the wind was able to satisfactorily explain the centroid positions in \cite{baezrubio2013}. In the model, the wind and the jet are launched from a certain radius at their expansion velocity \mbox{$v_{wind}=60$ \unit{km.s^{-1}}} and \mbox{$v_{jet}=250$ \unit{km.s^{-1}}}, respectively (Table \ref{tab:inputparameters}), with no wind or jet expansion inside that radius. Therefore, the launching mechanism is not accurately described, and it is introduced as a discontinuity between the disk and the wind. Besides, the velocity of the jet is not constant along the double cone due to the collimation (Sect. \ref{sec:jet}). The model describes the kinematics of the jet and the wind, but not the dynamical interaction that leads to the collimation of the jet, nor its interaction with the ionized wind. In this complex scenario, we have used a half-Keplerian rotation for the wind and the jet to describe the centroid map, which is an average between the rotating and non-rotating cases. A future version of the model with a more accurate description of the launching and collimation mechanisms will allow to elucidate the origin of this loss of efficiency in extracting angular momentum from the disk. Nevertheless, even if the wind is not co-rotating with the disk, the detection of rotation implies that it is removing angular momentum from it, as expected if accretion is taking place \citep{zhang2017}: we find that the ionized wind extracts $\sim$6 times more angular momentum than the ionized jet (Sect. \ref{sec:massmomentum}).
        
    \subsection{Deceleration of the radial expansion} \label{sec:bvdiscussion}
    
        As explained in Sect. \ref{sec:centroidmap}, a small radial deceleration exponent $b_v$ (see Sect. \ref{sec:bv}) has been introduced in the last step of the modeling to match the predicted high-velocity RRL emission with that of the observations. \cite{tafoya2021} find a radial deceleration in the high velocity component of the outflow ejected from the \mbox{$\sim$ 10 M$_{\odot}$} ALMA1 core in the \mbox{70 $\mu$m} dark clump \mbox{G010.991--00.082}. They offer two possible explanations. One of them proposes that the jet is expanding with a low Mach number ($<$ 6) and it is thus decelerating as it interacts with the ambient material \citep[see also][]{chernin1994}. The second possibility is that the jet is not being ejected at a constant velocity: the material located further from the star, which was ejected earlier, could have done it with a lower velocity than the more recent, subsequent ejections. The jet velocity we find in our model is $\sim$10 times faster than the sound speed in the \mbox{12000 K} ionized gas of MWC 349A \citep[][Table \ref{tab:inputparameters}]{baezrubio2013}. CO absorption has been detected in observations towards the star \citep[e.g.][]{martinpintado1994,mitchell1990}, and CO emission is seen in the vicinity at the radial velocity of MWC 349A \citep{strelnitski2013}. This could imply that the star is embedded in a molecular cloud and that the wind and the cloud may be interacting. The explanation for this slight deceleration still remains speculative and additional observations of the jet, probing larger distances, would be needed to explain this finding. Besides, episodic ejections, as considered in Sect. \ref{sec:jetdiscussion} could complicate the picture. 
        
    \subsection{Saturation and extent of the H26$\alpha$ maser emission} \label{sec:h26adiscussion}
    
        As presented in Section \ref{sec:h26a}, the H26$\alpha$ maser emission from MWC 349A is saturated \citep{baezrubio2013,tran2021}, and therefore we have considered the saturation effect when modeling the ALMA observations of the H26$\alpha$ line. The treatment of the saturation in the MORELI code is, however, an approximation in the sense that saturation of the maser is only considered along each line of sight. More accurate treatments, instead, consider the ray-tracing algorithm in which photons coming from any direction contribute to the saturation of the maser \citep[see][]{tran2021}. Still, the modeled H26$\alpha$ centroid map and line profile agree qualitatively with the ALMA observations and serve as an additional confirmation of the newly discovered jet component. The largest deviation of the model is at the extremes of the disk branch, where the maser amplification is stronger, and hence the H26$\alpha$ line emission is more saturated. A better match in the centroid map may be obtained with a refinement of the saturation treatment in the MORELI code.
        
        The maser peaks arise from a nearly edge-on disk at a certain radius that has the optimum electron density for the amplification of each recombination line \citep{strelnitski1996,weintroub2008,zhang2017}. The extent of the disk branch in the H26$\alpha$ centroid map is smaller than in H30$\alpha$, as it requires higher densities for the amplification; this was already noticed by \cite{zhang2017} in their SMA observations of the two RRLs, which they found to be consistent with the maser theory assuming that the density of the ionized gas decreases with the radius.

\section{Conclusions} \label{sec:conclusions}

    In this work we have modeled the H30$\alpha$ and H26$\alpha$ maser-amplified RRL emission recently measured towards the massive star MWC 349A with ALMA by \cite{prasad2023}. The observations allow us to perform spectro-astrometry on the lines with a relative photocenter accuracy of \mbox{$\sim$1 mas}, i.e. \mbox{$\sim$1.2 au}, which is achieved thanks to the strong maser amplification of the RRLs and allows for a detailed study of the kinematics of the innermost ionized regions of the star.
    
    In order to explain the RRL features detected by ALMA, we have expanded the kinematical options of the radiative transfer code MORELI \citep{baezrubio2013}, previously used to describe the ionized gas in MWC 349A. From our modeling, we infer that the presence of a high-velocity jet inside the ionized wind can reproduce the higher radial velocity RRL emission unveiled by ALMA. The bipolar jet in our best-fit model is expanding at a velocity of 250 \unit{km.s^{-1}} and it is
    tilted $\sim$16--22$^{\circ}$ with respect to the rotation axis of the disk, with the northern jet facing the observer, and the southern one expanding away. We also find a low collimation degree for the jet: the velocity is reduced to half the maximum value at a semi-opening angle of 24$^{\circ}$. The jet is coupled with the previously detected wide-angle disk wind that expands radially at 60 \unit{km.s^{-1}}. In our model, both components are rotating with half of the Keplerian rotation from the disk, and the radial expansion of the jet and wind is slightly decelerated.
    
    Based on these results, we draw the following implications for our understanding of massive star formation: 
    
    \begin{itemize}
        \item{The jet in MWC 349A is consistent with the magnetohydrodynamic disk wind model \citep{koniglpudritz2000}, which has previously been proposed for this star \citep{martinpintado2011,baezrubio2014,zhang2017}. The derived mass loss and momentum rates are consistent with those of the well studied massive system \mbox{Cep A HW2}. Since this is the first detection of the high velocity emission, a sudden ejection of material cannot be discarded; future observations will shed light on a possible episodic or variable nature of the ejection.}
        
        \item{The tilt of the jet could be caused by an inner unresolved stellar companion, whose presence has been proposed before \citep{jorgenson2000,hofmann2002,gvaramadzementen2012}. A warped disk as in Mon R2-IRS2 \citep{jimenezserra2020}, which may be produced by the presence of an inner companion, could also explain the deviation of the jet expansion with respect to the rotation axis.}
        
        \item{The jet is not well collimated: if MWC 349A is a young star in an advanced stage of its formation, a wide-angle jet would be consistent with the evolutionary sequence proposed by \cite{beuthershepherd2005} and the observational findings of \cite{frost2021}, where cavities in young massive stars become wider with age.}
        
        \item{Rotation of the outflow implies removal of angular momentum from an accreting disk, as described in \cite{zhang2017}. The jet in our model rotates in the same sense as the disk, which supports the scenario of the star still accreting material from its circumstellar disk. We use a half-Keplerian rotation for the wind and jet as an average between the rotating and non-rotating cases. This will possibly be explained by a model with a detailed description of the outflow launching and collimation, as well as the interaction between jet and wind. We find that the ionized jet is extracting $\sim$6 times less angular momentum than the ionized wind has extracted, and that they combined have extracted about 25\% of the angular momentum of the disk}.
        
        \item{An adequate treatment of the saturation effects of the maser in the H26$\alpha$ emission is necessary to accurately reproduce the observations. The line is strongly saturated according to the predictions of the best model obtained in this paper, in contrast to H30$\alpha$ (\cite{baezrubio2013} and \cite{tran2021}), and the emission occurs closer to the central star in agreement with \cite{zhang2017}. This is the expected result since the line requires a higher density for the amplification, which is found nearer to the star.}  
        
    \end{itemize}

    This research has been funded by grant No. PID2019-105552RB-C41 by the Spanish Ministry of Science and Innovation/State Agency of Research MCIN/AEI/10.13039/501100011033. A.M.-H. has received support from grant MDM-2017-0737 Unidad de Excelencia "María de Maeztu" Centro de Astrobiología (CAB, CSIC-INTA) funded by MCIN/AEI/10.13039/501100011033. N.H. is funded by Spanish MCIN/AEI/10.13039/501100011033 grant PID2019-107061GB-C61. This paper uses the following ALMA data: ADS/JAO. ALMA No. 2017.1.00404.S. ALMA is a partnership of ESO (representing its member states), NSF (USA) and NINS (Japan), together with NRC (Canada), MOST and ASIAA (Taiwan), and KASI (Republic of Korea), in cooperation with the Republic of Chile. The Joint ALMA Observatory is operated by ESO, AUI/NRAO, and NAOJ. The National Radio Astronomy Observatory is a facility of the National Science Foundation operated under cooperative agreement by Associated Universities, Inc.

\appendix

\section{Appendix figures}

    \begin{figure}[h]
        \includegraphics[width=18cm, height=9cm]{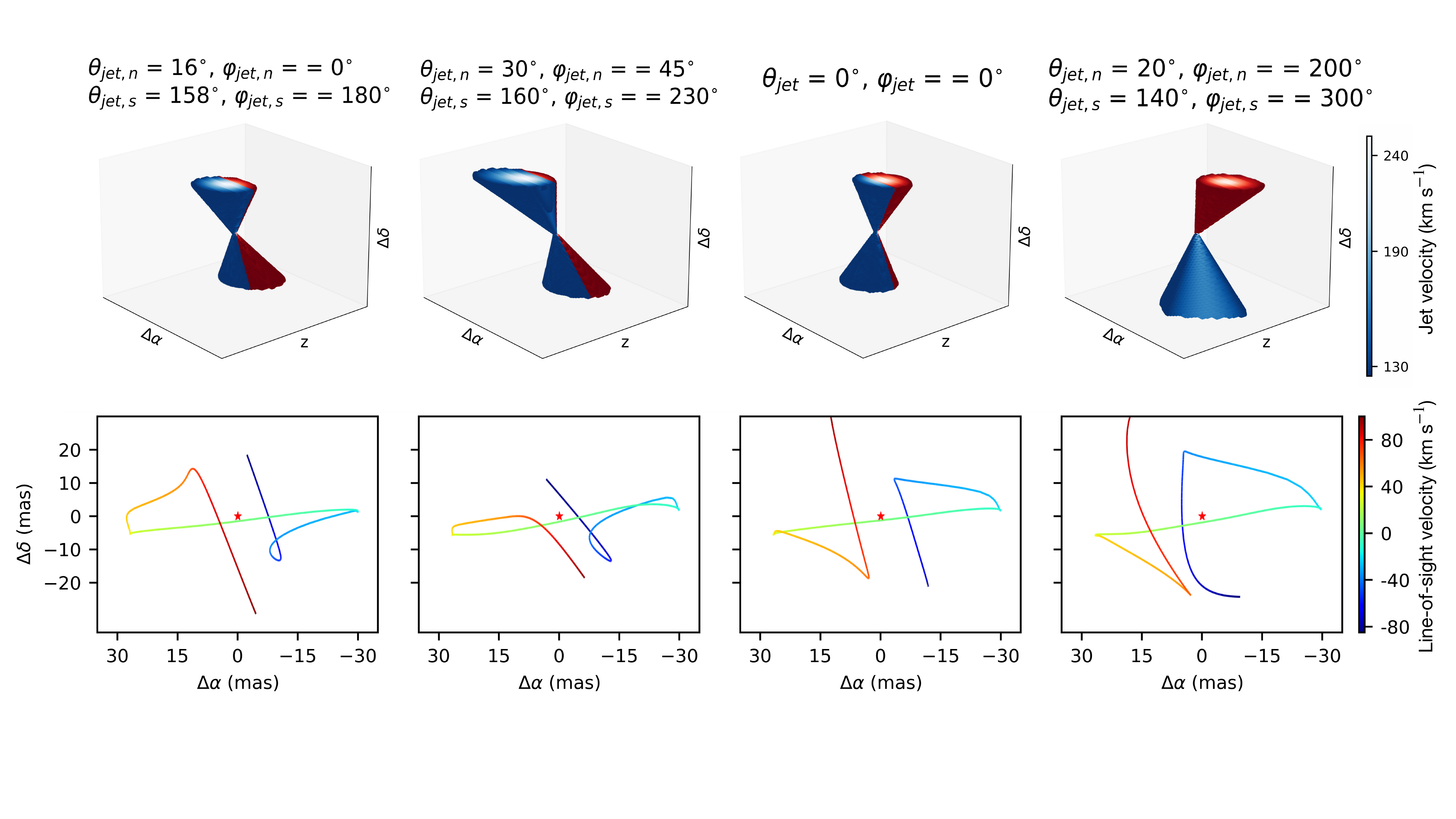}
        \centering
        \caption{Four different jet orientations (upper panels) and the resulting H30$\alpha$ centroid maps from the model (lower panels). The jet panels represent those points where the jet velocity is $v\geq0.5v_{jet}$ for a collimation $\psi_0=0.4$. This corresponds to a semiopening angle of 24$^{\circ}$ by substituting in Equation \ref{eq:psiweight}. Blue points are those where $z<0$, i.e. where the emission is blueshifted as seen by the observer; red points, conversely, represent redshifted emission. The jet orientation $(\theta_{jet},\varphi_{jet})$ for each case is indicated at the top of the upper panels. The leftmost model corresponds to the final model presented in Figure \ref{fig:panelcentroids}. \label{fig:jetorientations}}
    \end{figure}
    
    \begin{figure}[h]
        \includegraphics[width=18cm, height=5.5cm]{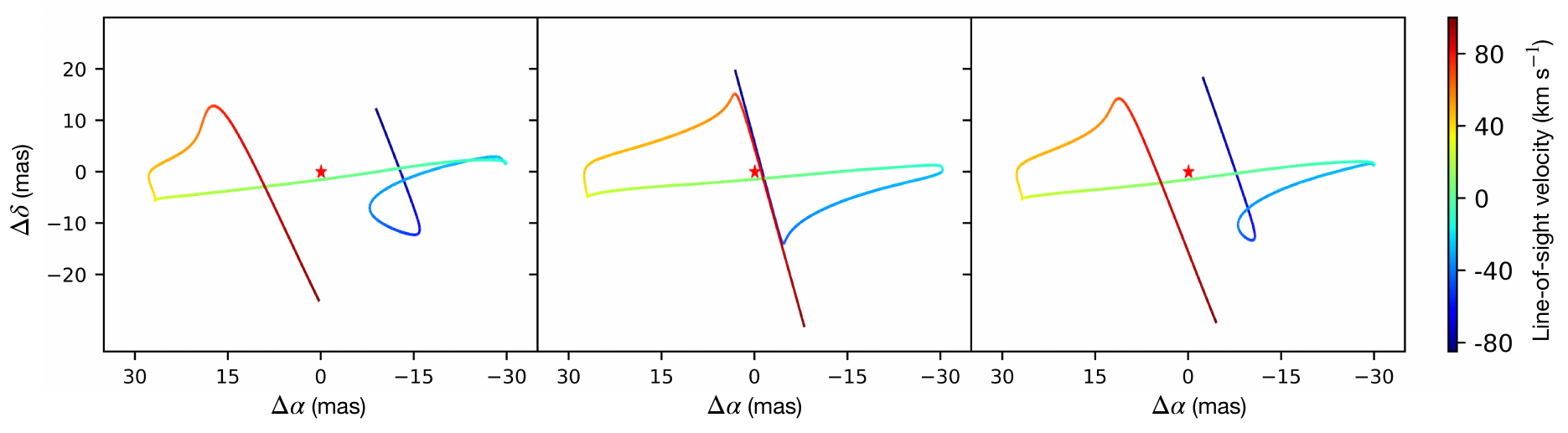}
        \centering
        \caption{Centroid maps of the H30$\alpha$ line from the same model with full Keplerian rotation of the outflow (left), no rotation (center) and half of the rotation (right), which is the final model. The more rotation included in the model, the further away the jet branches get from the rotation axis. \label{fig:centroidsrot}}
    \end{figure}
    
    \begin{figure}[ht]
        \includegraphics[width=18cm, height=7.5cm]{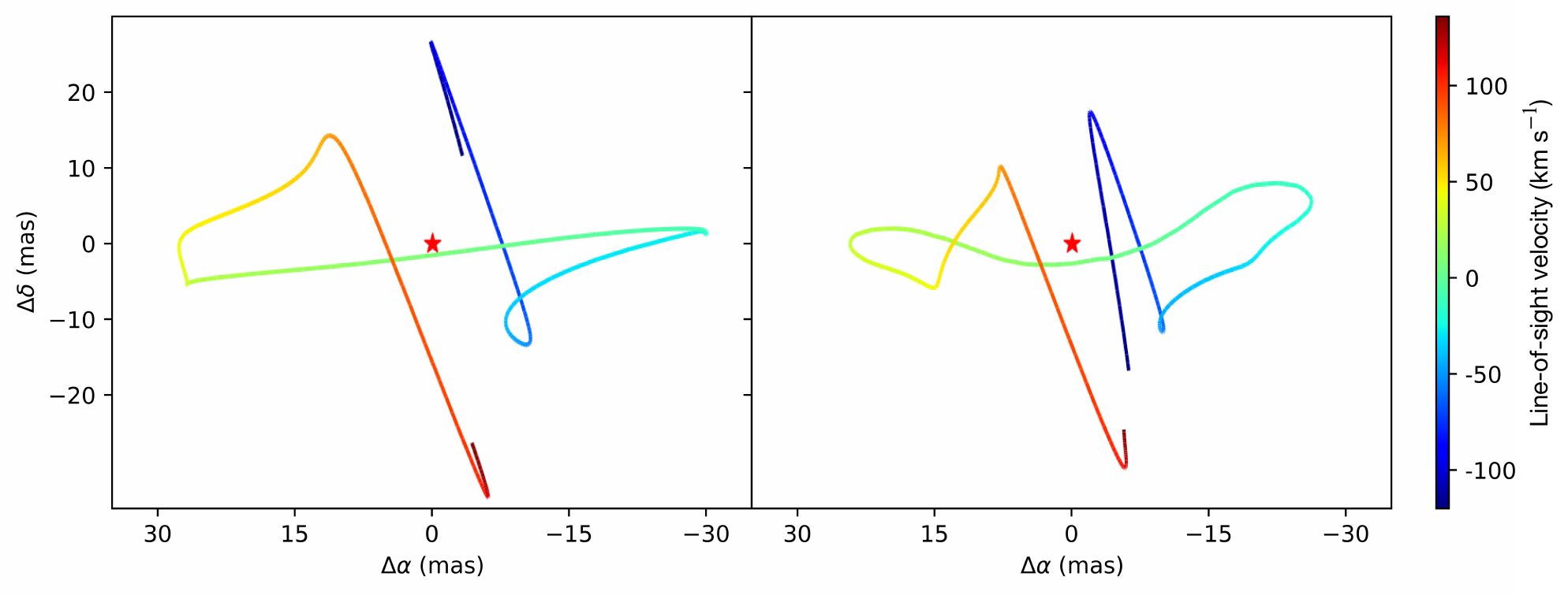}
        \centering
        \caption{Centroid maps for the H30$\alpha$ (left) and H26$\alpha$ (right) lines from the model with an extended radial velocity range of (-120, 135) \unit{km.s^{-1}} to show the behaviour of the second turnaround of the jet branch at the most extreme radial velocities. The shift towards the disk seen in the observations is predicted by our best-fit model, but at a larger velocities than observed with ALMA as shown here.}
        \label{fig:centroidsfullrange}
    \end{figure}

    \begin{figure}[ht]
        \includegraphics[width=19cm, height=5.5cm]{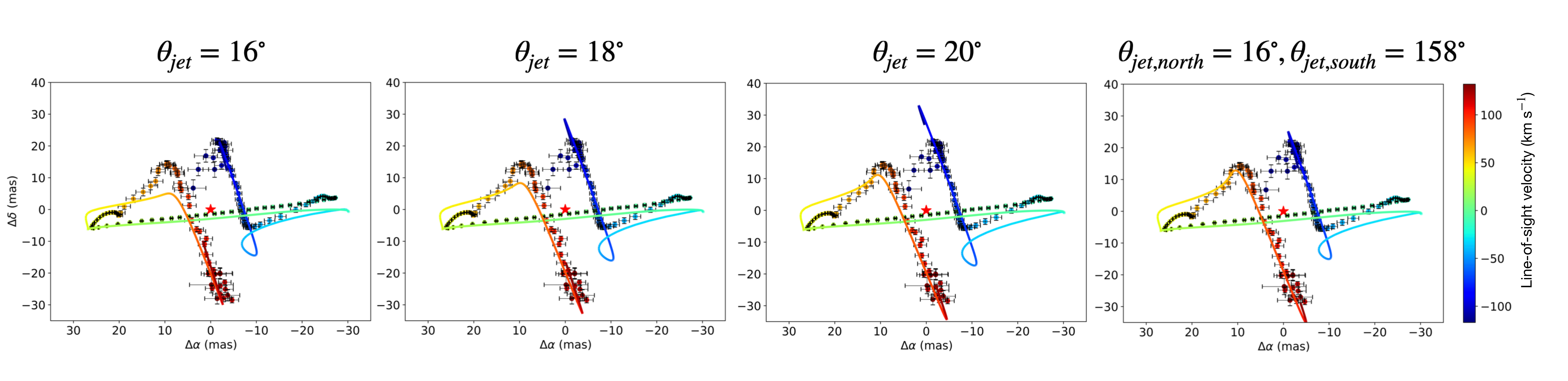}
        \centering
        \caption{Centroid maps for the H30$\alpha$ emission from observations (colored points with errors) and from the model (color line) with an extended radial velocity range of (-120, 135) \unit{km.s^{-1}} to show the effect of the symmetrical orientation of the jet versus a slight asymmetry. The three leftmost panels represent a symmetrical jet, where $\theta_{jet}$ represents the orientation of the northern jet facing the observer ($\varphi_{jet}=0^{\circ}$). The rightmost panel corresponds to the final model presented in Figure \ref{fig:panelcentroids} consisting of a slightly asymmetrical jet, where the southern cone is more inclined than the northern. In the four models, $\varphi_{jet, north}=0^{\circ}, \varphi_{jet, south}=180^{\circ}$.}
        \label{fig:asymmetricaljet}
    \end{figure}
    
    \begin{figure}[ht]
        \includegraphics[width=13cm, height=7.5cm]{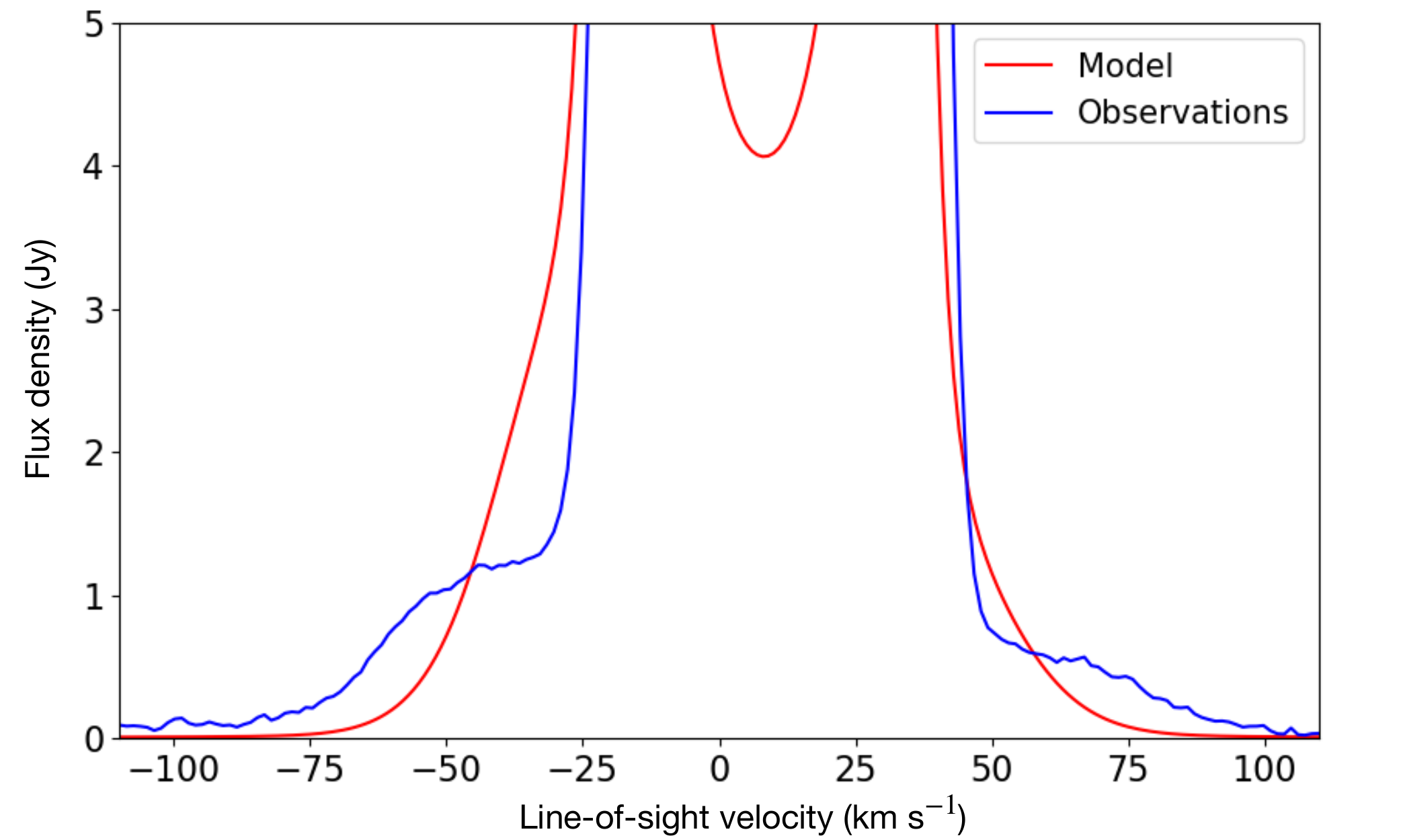}
        \centering
        \caption{H30$\alpha$ line profile from observations (blue) and a model (red) where only the ionized wind expanding at 60 \unit{km.s^{-1}} is included. We show a zoom in at the lower intensity values to highlight the behaviour of the wing humps.}
        \label{fig:h30alinewithoutjet}
    \end{figure}

\clearpage
\bibliography{sample631}{}
\bibliographystyle{aasjournal}



\end{document}